\documentclass[12pt]{article}%
\usepackage{amssymb}
\usepackage{amsmath}
\usepackage{amsfonts}
\usepackage{graphicx,wrapfig,lipsum}
\providecommand{\U}[1]{\protect\rule{.1in}{.1in}}

\usepackage{tikz}
\usetikzlibrary{decorations.pathmorphing}
\usetikzlibrary{math}
\usetikzlibrary{fadings}

\begin{document}
\bigskip%

\begin{titlepage}
\vspace{.3cm} \vspace{1cm}
\begin{center}
\baselineskip=16pt \centerline{\bf{\Large{Non-flat Universes and Black Holes }}} 
\baselineskip=24pt \centerline{\bf{\Large{ in Asymptotically Free Mimetic Gravity  }}}
\vspace{1truecm}

			\centerline{\large\bf Ali H.
				Chamseddine$^{1,2}$\ , Viatcheslav Mukhanov$^{2,3,4}$\ , Tobias B. Russ$^{2}%
				$\ \ } \vspace{.5truecm}
			\emph{\centerline{$^{1}%
					$Physics Department, American University of Beirut, Lebanon}}
			\emph{\centerline{$^{2}%
					$Theoretical Physics, Ludwig Maxmillians University,Theresienstr. 37, 80333 Munich, Germany }%
			}
			\emph{\centerline{$^{3}%
					$MPI for Physics, Foehringer Ring, 6, 80850, Munich, Germany}}
			\emph{\centerline{$^{4}%
					$School of Physics, Korea Institute for Advanced Study, Seoul 02455, Korea}}
		\end{center}
\vspace{2cm}
\begin{center}
{\bf Abstract}
\end{center}
The recently proposed theory of ``Asymptotically Free Mimetic Gravity'' is extended to the general non-homogeneous, spatially non-flat case. We present a modified theory of gravity which is free of higher derivatives of the metric. In this theory asymptotic freedom of gravity implies the existence of a minimal black hole with vanishing Hawking temperature. Introducing a spatial curvature dependent potential, we moreover obtain non-singular, bouncing modifications of spatially non-flat Friedmann and Bianchi universes.
\end{titlepage}%
\tableofcontents

\clearpage
\section{The Theory}
\subsection{Introduction}
General relativity is a quintessential example of a successful physical theory. At the same time as its conceptional simplicity is striking, it has to this day not failed a single experimental test.
However, one theoretical prediction of GR that contradicts physical intuition is the formation of singularities from  physically realistic initial conditions, as was shown in the singularity theorems of Hawking and Penrose \cite{HawkingPenrose}. The standard approach to argue away these singularities is to delegate their removal to a hypothetical theory of quantum gravity which should begin to take over as the curvature approaches its Planckian value. For a variety of reasons it is clear that GR is bound to fail at this scale. Moreover, at this point even the description of our physical world as a smooth manifold is no longer justified, complicating drastically any approach to quantum gravity.

A different approach to singularity resolution is to allow deviations from GR already at curvature scales some orders below the Planck curvature where the smooth manifold description of spacetime is still a sensible concept. If we can manage to modify gravity at these scales in such a way as to implement an upper bound on all curvatures, we could get rid of singularities on a classical level. Moreover, we would never enter the Planck regime and thus potentially avoid the practical need for a theory of quantum gravity altogether.

Einstein gravity is distinguished among all local, covariant theories of metric gravity in four spacetime dimensions by the fact that it has equations of motion which are only second order. It seems that the only chance to alter this theory and not end up with higher derivatives is to bring something else other than the metric into the game.
In the case of the mimetic field, however, this is not an entirely new physical entity. Rather, it represents a reshuffling of the degrees of freedom of the metric itself. The starting point of  so-called ``mimetic gravity'' was in \cite{mimetic} to reparametrize the physical metric $g_{\mu\nu}$ in the form
\begin{equation}
g_{\mu\nu}=h_{\mu\nu}h^{\alpha\beta}\phi_{,\alpha}\phi_{,\beta} %
\end{equation}
in terms of an auxiliary metric $h_{\mu\nu}$ and a scalar field $\phi$, called the mimetic field.
Since the physical metric is invariant under Weyl transformations of $h_{\mu\nu}$, the mimetic field takes over the job of representing the conformal degree of freedom of gravity. 
By definition,  $\phi$ identically satisfies
\begin{equation}
g^{\mu\nu}\phi_{,\mu}\phi_{,\nu}=1, \label{constraint}%
\end{equation}
and we can impose the nature of the mimetic field also by adding this condition as a constraint to the gravity action \cite{Golovnev}, giving it the general form
\begin{equation*}
S=\frac{1}{16\pi}\int\textup{d}^{4}x\sqrt{-g}\left(-\mathcal{L}\left[g_{\mu\nu},\phi\right]  +\lambda\left(  g^{\mu\nu}\phi_{,\mu}%
\phi_{,\nu}-1\right) \right),   \label{3}%
\end{equation*}
where $\lambda$ is a Lagrange multiplier. This formulation is advantageous over simply applying the reparametrization because at the end of the day we are interested in a theory of the physical metric rather than a theory of the auxiliary metric. Note that constraint (\ref{constraint}) can be derived as a consequence of 3d volume quantization in noncommutative geometry \cite{Quanta}, \cite{Hilbert}.

Inserting the Einstein-Hilbert Lagrangian $\mathcal{L} = R[g_{\mu\nu}]$ just reproduces standard GR with an additional contribution of mimetic matter (cf.  \cite{mimetic}, \cite{Golovnev}). Different choices for $\mathcal{L}$ with added potentials depending on $\phi$ or $\Box \phi$  have been considered in \cite{mimcos}, \cite{Singular}, \cite{BH} and the mimetic field has also been used successfully to define ghost-free massive gravity \cite{massivmim1}, \cite{massivmim2}.

Recently, in \cite{AFmimetic}, it was realized that the mimetic field can be used to implement in a covariant manner the idea of a running gravitational and cosmological constant by means of a Lagrangian of the form
\begin{equation}
     \mathcal{L} = f[\phi] R[g_{\mu\nu}] + 2\Lambda[\phi]
\end{equation}
where the ``inverse gravitational constant'' $f$ and ``cosmological constant'' $\Lambda$ can depend on $\phi$ and its derivatives in a way to be determined. To this end, let us find out which covariant quantities can be constructed from $\phi$. First, note that by virtue of (\ref{constraint}), $t:=\phi$ qualifies to be used as the time coordinate of a synchronous coordinate system (cf. appendix A),
\begin{equation}
\textup{d}s^{2}=\textup{d}t^{2}-\gamma_{ab}\textup{d}x^{a}\textup{d}x^{b}.
\label{synch}%
\end{equation}
Hence, a simple $\phi$ dependence of $f$ and $\Lambda$ would resemble the introduction of a time dependent background. Moreover, the only covariant quantity constructed from first derivatives of $\phi$ is identically constant by (\ref{constraint}). Gratifyingly, the second covariant derivatives of $\phi$, however, represent measures of the curvature related to the conformal degree of freedom of the gravitational field. More precisely,
\begin{equation*}
    -\phi_{;ab} = \kappa_{ab} = \frac{1}{2}\frac{\partial}{\partial t}\gamma_{ab} 
\end{equation*}
is the extrinsic curvature of the slices of constant $\phi$, while $\phi_{;0\alpha}=0$.
The Ricci scalar written out in this synchronous slicing given by $\phi$ reads
\begin{equation}
    -R = 2\dot{\kappa}+\kappa^2+\kappa_{b}^{a}\kappa_{a}^{b}+ {^3\!R},
    \label{Ricciscalar}
\end{equation}
where dot denotes $t$-derivatives, $\kappa_{b}^{a}=\gamma^{ac}\kappa_{cb}$, ${^3\!R}$ is the 3-curvature of the spatial slices and
\begin{equation*}
\kappa : = \gamma^{ab}\kappa_{ab}  = g^{\alpha\beta}\phi_{;\alpha\beta} = \Box\phi%
\end{equation*}
is the trace of extrinsic curvature. From expression (\ref{Ricciscalar}) we can read off the reason why the Einstein equation is only second order: because second derivatives of the metric appear only linearly in $R$ and thus only contribute as total derivatives to the action. The only chance to introduce a curvature dependence of the gravitational constant and not spoil this property is $f[\phi] = f(\Box\phi)$.
In this case
\begin{equation}
   -f(\Box \phi)R=2 \dot{F}(\kappa) + f(\kappa)\left(\kappa^{2}+\kappa_{b}^{a}\kappa_{a}^{b}+{^{3}\!R}\right) 
\label{5}
\end{equation}
where $f$ is assumed to be integrable with $f(\kappa)=F^{\prime}(\kappa)\equiv\partial F / \partial \kappa$. Since up to a total derivative such a Lagrangian still contains only first time derivatives of the metric, we can expect the modified Einstein equation of such a theory to be second order in time. While the arguments used to arrive at this result were rather heuristic\footnote{Strictly speaking, it is not
allowed to use $\Box\phi=\kappa$ and impose gauge conditions in the action
before variation.}, we can of course explicitly verify this statement. 
Indeed, the theory defined by
\begin{equation}
    \mathcal{L}=  f(\Box\phi) R +  2\Lambda(\Box\phi)
    \label{6}
\end{equation}
that has been studied in \cite{AFmimetic} turned out to be free of higher time derivatives in the synchronous frame. In the general spatially non-flat case, however,  higher spatial and mixed derivatives will appear. The origin of these terms can be traced back to the presence of $f(\Box \phi)$ in front of ${^{3}\!R}$ in (\ref{5}). Luckily, we can use $\phi$ to dissect this term in a covariant way as
\begin{equation*}
   \widetilde{R}= 2\phi^{,\mu}\phi^{,\nu}G_{\mu\nu} - (\Box \phi)^2 + \phi ^{;\mu\nu}\phi_{;\mu\nu}  \overset{\cdot}{=} {^3\!R},
\end{equation*}
where $G_{\mu\nu}$ is the Einstein tensor and $\overset{\cdot}{=}$ means equality under the condition that (\ref{constraint}) is satisfied.
Subtracting the term that was added involuntarily, we will find the theory defined by
\begin{equation*}
    \mathcal{L} = f(\Box\phi)R+ (f(\Box\phi)-1) \widetilde{R} +2\Lambda(\Box\phi)
    \label{eq-Lgrav}
\end{equation*}
to be free of higher derivatives of all sorts. The addition of the second summand is hence motivated by the same argument that made Einstein gravity unique. 

For generality, we will still find it useful to include also a non-linear, spatial curvature dependent potential $h(\widetilde{R})$. It is clear that thereby higher spatial derivatives will reappear,  but no higher time or mixed derivatives. While higher time derivatives would typically introduce additional (potentially ghost-like) degrees of freedom, higher spatial derivatives could actually be useful to improve the renormalizability of gravity, along the lines of Ho\v{r}ava gravity \cite{MimeticHorava}. 

For this extended Lagrangian the interpretation of the functions $f$ and $\Lambda$ in terms of gravitational and cosmological constant might look less transparent than for (\ref{6}). However, we would like to stress that the homogeneous, spatially flat sectors of both theories are identical.

In the context of a theory like (\ref{6}) it is natural to realize the concept of limiting curvature (cf.  \cite{Markov},\cite{MukBran},\cite{Mukbransol}) by limiting the measure of curvature provided by $\Box \phi$.
Motivated by the analysis of the anisotropic sector made in \cite{AFmimetic}, the concept of ``asymptotic freedom'' of gravity gains special importance. This name is awarded to modifications where such a limiting curvature is implemented by a vanishing of the gravitational constant at some limiting value $ \Box\phi =\kappa_0$ which is a free parameter of the theory and can be chosen well below the Planckian curvature.

In this paper, after presenting the general form of the theory motivated above, we will consider applications to spatially non-flat Friedmann and Bianchi universes and black holes, providing more detailed calculations for the results presented in \cite{BlackHoleRemnants}.

\newpage
\subsection{Action and equations of motion}
Let us consider the theory defined by the action
\begin{equation}
S=\frac{1}{16\pi}\int\textup{d}^{4}x\sqrt{-g}\left(-\mathcal{L}  +\lambda\left(  g^{\mu\nu}\phi_{,\mu}%
\phi_{,\nu}-1\right)  \right)  , \label{action}%
\end{equation}
where%
\begin{equation}
\mathcal{L} = f(\Box\phi)R+ (f(\Box\phi)-1) \widetilde{R}  +2\Lambda(\Box\phi) + h(\widetilde{R} )
\label{Lgrav}%
\end{equation}
and 
\begin{equation}
   \widetilde{R}= 2\phi^{,\mu}\phi^{,\nu}G_{\mu\nu} - (\Box \phi)^2 + \phi ^{;\mu\nu}\phi_{;\mu\nu}. \label{7}
\end{equation}
This action contains two free functions $f$ and $\Lambda$ of $\Box \phi$, representing the inverse running gravitational constant $G(\Box\phi)^{-1}$ and cosmological constant\footnote{Note that for this interpretation we still need to factor out the gravitational constant such that $\Lambda = f \overline{\Lambda}.$} $\bar{\Lambda}(\Box\phi)$, respectively.
For generality, we also included a spatial curvature dependent potential $h(\widetilde{R})$. 
In the following we will use Planck units setting $ G\left(
\Box\phi=0\right)  =G_0=1$, such that $f\left(  \Box\phi=0\right)=1.$

Variation of the action with respect to the metric $g_{\mu\nu}$ yields the modified Einstein equation

\begin{align}
(1&-h')R_{\mu\nu}-\left (\tfrac{1}{2}\mathcal{L}+(\Tilde{Z}\phi^{,\alpha} )_{;\alpha} +\Box h'  \right )g_{\mu\nu} +\left (\phi_{,\mu}\phi_{,\nu}    \Tilde{f}^{,\alpha} -\phi_{;\mu\nu} \Tilde{f} \phi^{,\alpha} \right )_{;\alpha} 
\phantom{\frac{1}{1}} \nonumber \\
 &+ 2\Tilde{f}\phi^{,\alpha}\phi_{(,\mu}R_{\nu)\alpha}+  2 \phi_{(,\mu} \Tilde{Z}_{,\nu )} +h'_{;\mu\nu}
= (\lambda + \Tilde{f}R) \phi_{,\mu}\phi_{,\nu} + 8\pi T_{\mu\nu}^{\textup{(m)}}, \phantom{\frac{1}{1}}
\label{mEE}
\end{align}

where
\begin{equation*}
  \Tilde{f} := f-1+h',\qquad Z:=\tfrac{1}{2}f'\left((\Box\phi)^2+\phi^{;\mu\nu}\phi_{;\mu\nu} \right) - \Lambda', \qquad \Tilde{Z} := Z - \phi^{,\alpha}h'_{,\alpha},
\end{equation*}
$f':= \mathrm{d}f/\mathrm{d}\Box\phi$, $\Lambda':= \mathrm{d}\Lambda/\mathrm{d}\Box\phi$, $h':= \mathrm{d}h/\mathrm{d} \widetilde{R}$ and $T_{\mu\nu}^{\textup{(m)}} = \tfrac{2}{\sqrt{-g}}\tfrac{\delta S^\textup{m}}{\delta g^{\mu\nu}} $ is the matter energy momentum tensor. For the detailed calculations in the variation of the action the interested reader is referred to appendix B.
While at first glance this modified Einstein equation looks more involved than the one presented in \cite{AFmimetic}, calculating its components in the synchronous frame will reveal that in the general case, apart from the new terms due to $h$, the theory defined by (\ref{Lgrav}) is in fact the simpler one. 

\clearpage
The evolution of the mimetic field is already completely determined by the constraint (\ref{constraint}), which we obtain from variation with respect to the Lagrange multiplier.
The equation obtained by varying (\ref{action}) with respect to $\phi$ hence can only return the favor and provide a condition to determine $\lambda$.
Conveniently written in terms of the quantity 
\begin{equation}
    \Xi := \lambda +\Tilde{f}\left(R-R_{\mu\nu}\phi^{,\mu}\phi^{,\nu}\right)-\Box {f} -\phi^{,\mu}{Z}_{,\mu}- \phi^{,\mu}{h'}_{,\mu} \Box \phi,
    \label{Xi}
\end{equation}
this ``equation of motion'' of $\phi$ reads

\begin{align}
    \left(\Xi \,\phi^{,\nu}\right)_{;\nu}=  \left [  (f-h')^{,\mu}\phi^{;\nu}_{\mu} + Z^{,\nu}-\phi^{,\nu}\phi^{,\mu}Z_{,\mu}+\Box\phi \left(h'^{,\nu}-\phi^{,\nu}\phi^{,\mu}h'_{,\mu}\right)  \phantom{\frac{1}{1}}\right. \nonumber\\ + \left. \phantom{\frac{1}{1}}\Tilde{f}\left ( R^{\mu\nu}\phi_{,\mu}- R^{\alpha\beta}\phi_{,\alpha}\phi_{,\beta}\phi^{,\nu} \right ) \right ]_{;\nu}.
\end{align}
In the synchronous frame the right hand side turns out to be just the 3-divergence of a 3-vector (denoted by $ X^a_{|a}$) and we find the solution
\begin{equation}
    \Xi = \tfrac{1}{\sqrt{\gamma}} \int \!\textup{d}t \sqrt{\gamma}\, \left( \kappa^a_{b}f^{,b}+Z^{,a} +\Tilde{f}R^{a}_{0} +\kappa h'^{,a}-\kappa^a_{b}h'^{,b} \right)_{|a}.
    \label{Xisol}
\end{equation}

Let us now evaluate (\ref{mEE}) in the synchronous frame $t=\phi$ where it takes its simplest form. 
Using (\ref{Xi}), the $0-0$ component of the modified Einstein equation becomes
\begin{equation}
    \frac{1}{3}\left(  f-2\kappa f^{\prime}\right)  \kappa^{2}-\Lambda
+\kappa\Lambda^{\prime}-\frac{1}{2}\left(  f+\kappa f^{\prime}\right)
\tilde{\kappa}_{b}^{a}\tilde{\kappa}_{a}^{b}= \frac{1}{2}\left(h-{^3\!R}\right) +  \Xi + 8\pi T^{\textup{(m)}}_{00},
\label{mEE00}
\end{equation}
where $\tilde{\kappa}_{b}^{a} : = \kappa_{b}^{a}-\tfrac{1}{3}\kappa \delta_{b}^{a}$ is the traceless part of the extrinsic curvature.
Note that inserting the solution (\ref{Xisol}) for $\Xi$, this equation becomes in general an integro-differential equation. Suitably taking another time derivative $\left(\% \phi^\mu\right)_{;\mu}$ of (\ref{mEE00}) yields a differential equation containing second time derivatives of the metric. This is a manifestation of the fact that in mimetic gravity the conformal degree of freedom of the gravitational field becomes dynamical.

Note that for spaces where ${^3\!R}=0$ and the integrand in $\Xi$ vanishes by homogeneity, (\ref{mEE00}) is precisely the same as the $0-0$ component of the modified Einstein equation presented in \cite{AFmimetic}.
\clearpage
The spatial components of (\ref{mEE}) after raising one index read
\begin{equation}
   -\frac{1}{\sqrt{\gamma}}\,\partial_t\left(\sqrt{\gamma}\,\left(f\kappa^{a}_{b}+Z \delta^{a}_{b}\right)\right) -\tfrac{1}{2}\mathcal{L}\,\delta^{a}_{b}  =   S^{a}_{b} +8\pi {T^{\textup{(m)}a}_{\phantom{\textup{(m)}}b}},
  \label{mEEab}
\end{equation}
where
\begin{equation}
    S^{a}_{b} : = (1-h') {^3\!R^{a}_{b}} + h'\,^{|a}_{\:b}- \Delta h'\delta^{a}_{b}
    \label{Sab}
\end{equation}
contains spatial curvature terms.
Subtracting from this equation one third of its (spatial) trace removes all isotropic terms proportional to $\delta^{a}_{b}$ and we obtain
\begin{equation}
   -\frac{1}{\sqrt{\gamma}}\,\partial_t\left(\sqrt{\gamma}\,f \,\tilde{\kappa}^{a}_{b}\right) = \tilde{S}^{a}_{b} +8\pi {\tilde{T}^{\textup{(m)}a}_{\phantom{\textup{(m)}}b}}
   \label{mEEabsub}
\end{equation}
where the right hand side consists of the traceless parts of $S^{a}_{b}$ and $T^{\textup{(m)}a}_{\phantom{\textup{(m)}}b}$.
The spatial components of the modified Einstein equation are hence second order in time. A non-linear potential $h({^3\!R})$ introduces higher spatial derivatives of up to forth order.

Finally, the mixed components of the modified Einstein equation (\ref{mEE}) read
\begin{equation}
    f R_{0a} +Z_{,a}+\kappa^b_{a}f_{,b} = 8\pi T_{0a}^{\textup{(m)}}.
    \label{mEE0a}
\end{equation}
These equations, as in standard GR, contain only first time derivatives of the metric and could be thought of as a constraint that needs to be satisfied on an initial hypersurface $\phi = \phi_{\textup{i}}$ and then continues to hold by virtue of validity of the other components of the modified Einstein equation.
Note that $h$ does not appear in the mixed equations. Moreover, (\ref{mEE0a}) can be used to simplify (\ref{Xisol}) to
\begin{equation}
    \Xi = \tfrac{1}{\sqrt{\gamma}} \int \!\textup{d}t \sqrt{\gamma}\, \left( T^{\textup{(m)}a}_{\phantom{\textup{(m)}}0}-(1-h')R^{a}_{0} +\kappa h'^{,a} - \kappa^{a}_{b} h'^{,b}\right)_{|a}.
    \label{Xisol2}
\end{equation}

Note that time reversal invariance of general relativity is maintained in our modification if we choose $f$ and $\Lambda$ as symmetric functions of $\kappa$. Moreover, if  we require
\begin{equation*}
    f=1+\mathcal{O}\left(  \kappa^{2}\right),\quad \Lambda=\mathcal{O}\left(  \kappa^{4}\right),\quad h = \mathcal{O}\left(\widetilde{R}^2\right) ,
\end{equation*}
then in the limit of low curvatures (\ref{mEE00}), (\ref{mEEab}) and (\ref{mEE0a}) are just the components of the usual Einstein equation with a contribution of mimetic matter, given by the constant of integration in $\Xi$.

\clearpage
\section{Non-flat Universes}
\subsection{Friedmann Universes}
The metric of a homogeneous, isotropic universe with cosmological time $t$ is given by
\begin{equation}
\textup{d}s^2=\textup{d}t^2-a^2(t)\left (\frac{\textup{d}r^2} {1-\varkappa r^2} +  r^2\textup{d}\Omega ^2\right ),
\end{equation}
where $\varkappa \in \left\{-1,0,+1 \right\}$ and $\mathrm{d}\Omega^2 = \mathrm{d}\vartheta^2+ \sin^2\vartheta\mathrm{d}\varphi^2$. Note that the unique solution of constraint (\ref{constraint}) in such a spacetime which is compatible with homogeneity is $\phi=t + const$.
Hence the quantities in the synchronous frame (\ref{synch}) are given by
\begin{equation*}
    \kappa = \frac{\dot{u}}{u} = 3 \,\frac{\dot{a}}{a},  \qquad \qquad {^3\!R}=\frac{6\varkappa}{a^2},
\end{equation*}
where we introduced $u:= a^3$. Moreover, by isotropy, $\tilde{\kappa}^{a}_{b}  = 0$ and by homogeneity  $\Xi \propto 1/u$ describes only the dust-like contribution of mimetic matter. 
For the sake of simplicity and because the remaining equation is still general enough for all our purposes we make again the simplifying choice
\begin{equation}
\Lambda=\frac{2}{3}\kappa^{2}(f-1), \label{18}%
\end{equation}
familiar from \cite{AFmimetic}. The $0-0$ modified Einstein equation (\ref{mEE00}) then becomes
\begin{equation}
    \left ( f-\frac{2}{3}\right ) \kappa^2= \frac{1}{2}\left(h({^3\!R})-{^3\!R}\right) + \varepsilon
    \label{mF}
\end{equation}
where $\varepsilon:= \Xi + 8\pi T^{(\textup{m})}_{00}$ is the total energy density of mimetic and ordinary matter. 
Note that this modified Friedmann equation is still formulated in terms of the same variables as the original Friedmann equation. Only the relation between curvature and energy density is changed at large curvatures. While the left hand side of (\ref{mF}) depends on $\kappa^2$ only, the right hand side is in general some function of $u$. Such a relation can be thought of as an integral curve in the ``phase space'' spanned by $u$ and $\kappa=\mathrm{d}\ln{u}/\mathrm{d}t$. Drawing this phase portrait for a specific modification will allow us to understand its qualitative behaviour without the need to obtain explicit solutions. 

\paragraph{Spatially flat universe:}
Let us use this phase portrait technique to show that in the case $\varkappa =0$ there is essentially only one possibility for the behaviour of a non-singular modification. Since the total energy density $\varepsilon$ is in general a monotonically decreasing function in $u$, in this case (\ref{mF}) can be understood as a relation of the form $u(\kappa^2)$. Furthermore, if this relation is to describe a non-singular modified Friedmann equation, it must be one-to-one. Otherwise, if at some point $\mathrm{d}u/\mathrm{d}\kappa^2 =0$, then
\begin{equation}
    \dot{\kappa} = \dot{u}\frac{\mathrm{d}\kappa}{\mathrm{d} u} =  u\,\kappa \frac{\mathrm{d} \kappa}{\mathrm{d} u} = \frac{u}{2}\,\frac{\mathrm{d} \kappa^2}{\mathrm{d} u}
    \label{kdot}
\end{equation}
would diverge. Since in the general non-vacuum case we cannot expect divergences of $\dot{\kappa}$ and $\kappa^2$ to cancel out exactly in the Ricci scalar (\ref{Ricciscalar}), both quantities have to be bounded separately to avoid a curvature singularity. Hence, in the following we will assume that the relation provided by (\ref{mF}) is of the form $\kappa^2(u)$ and it is one-to-one in the case $\varkappa=0$.

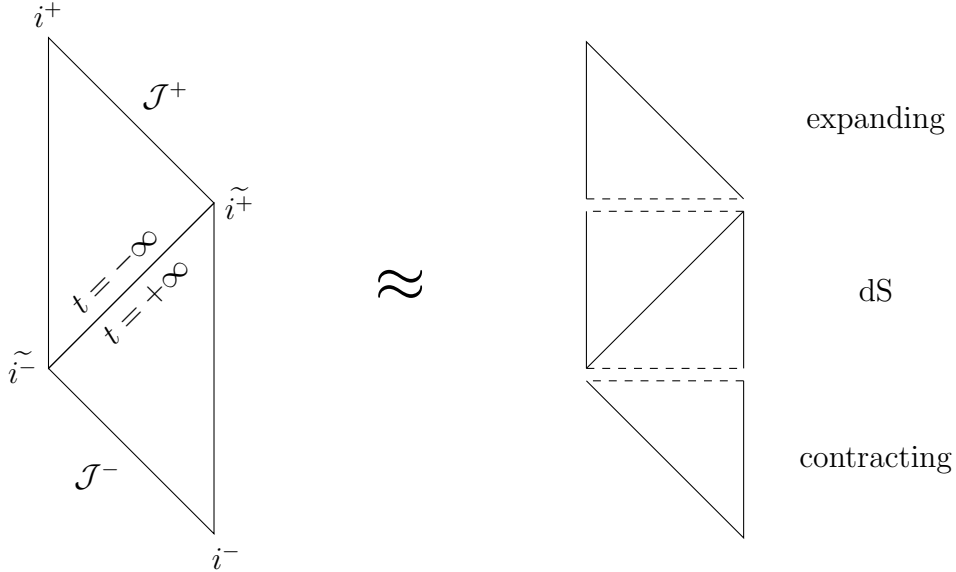
\begin{figure}[h]
    \centering
\begin{tikzpicture}[scale=0.55]
\tikzmath{\dx = 13;} 
\tikzmath{\s = 0.95;} 
\tikzmath{\dy = 6*(1-\s);}
\node (I)    at ( 4,0)   {};
\node (II)  at (0, 4) {};

\path (I) +(90:4)  coordinate[label=right:$\widetilde{i^+}$] (Itop)+(-90:4) coordinate[label=below:$\phantom{-}i^-$] (Ibot)+(180:4) coordinate (Ileft);
 
\path (II) +(90:4)  coordinate[label=above:$i^+$] (IItop)+(-90:4) coordinate[label= left:$\phantom{--}\widetilde{i^-}$] (IIbot);

\draw  (Ileft) --  node[midway, below left]{$\cal{J}^-$}(Ibot) -- (Itop) --node[midway, below, sloped]{$t=+\infty$} (Ileft);

\draw  (Itop) -- node[midway, above right]{$\cal{J}^+$}(IItop) -- (IIbot)--node[midway, above, sloped]{$t=-\infty$} (Itop);

\node (0) at ( 0.5*\dx+2,2) {\Huge{$\approx$}};

\node (0') at ( \dx+\s*2,\s*2)  {};
\node (0'tra) at ( \dx+\s*4,\s*4)  {};
\node (I')    at (\dx+\s*4,-\dy)   {};
\node (II')  at (\dx+0, \s*4+\dy) {};

\path (0'tra)+(0,0) coordinate(0'tr) +(-\s*4,0) coordinate(0'tl)+(-\s*4,-\s*4) coordinate(0'bl)+(0,-\s*4) coordinate(0'br);
\path (I')+(0:0) coordinate (I'right) +(0:-\s*4) coordinate (I'left)+(-90:\s*4) coordinate(I'bot);
\path (II')+(0:0) coordinate (II'left) +(0:-\s*4) +(0:\s*4) coordinate (II'right)+(90:\s*4) coordinate(II'top);

\draw[dashed]  (0'tr) -- (0'tl);
\draw (0'tl) --  (0'bl);
\draw[dashed] (0'bl) --(0'br);
\draw (0'br)--(0'tr);
\draw  (0'tr) -- (0'bl);
\draw [dashed] (I'right) -- (I'left);
\draw (I'left) --  (I'bot) -- (I'right);
\draw[dashed] (II'left) -- (II'right);
\draw (II'right) --  (II'top) -- (II'left);

\node (E) at (1.5*\dx+0.5,\s*6+\dy)  {expanding};
\node (D) at (1.5*\dx+0.5,\s*2)  {dS};
\node (C) at (1.5*\dx+0.5,-\s*2-\dy)  {contracting};

\end{tikzpicture}
    \caption{Conformal diagram of a modified spatially flat Friedmann universe.}
    \label{fig:CDfriedmann}
\end{figure}

By (\ref{kdot}), boundedness of $\dot{\kappa}$ implies that the domain of definition of the relation $\kappa^2(u)$ can be extended to $u\in\left[0,\infty\right)$. Boundedness of $\kappa$ implies that
\begin{equation}
    \int_{0}^{\infty}\!\mathrm{d}u \frac{\mathrm{d} \kappa}{\mathrm{d} u} = -\kappa(u=0)
\end{equation}
has a finite value, where we made use of the low curvature limit $\kappa^2(u\to\infty) = 0$. At the lower bound, the integral can only converge if $ u \frac{\mathrm{d} \kappa}{\mathrm{d} u} \to 0$. By (\ref{kdot}) it follows that $\dot{\kappa}\to 0$.
Hence, in this limit $\kappa$ must be asymptotically constant at some limiting value $\pm \kappa_0$. Recalling that $\kappa$ is the logarithmic derivative of $u$, this means that asymptotically
\begin{equation}
    u=a^3\propto \exp \left(\pm\kappa_0 t\right)
\end{equation}
as $t\to \mp \infty$. In conclusion, the most natural modifications generically replace Big Bang/Big Crunch singularities by a smooth transition to a de Sitter-like initial/final state with limiting curvature. 
In \cite{AFmimetic} we provided a concrete example of a non-singular, spatially flat universe using the simple choice
\begin{equation}
f=\frac{1}{1-\left(  \kappa^{2}/\kappa_{0}^{2}\right)  }. \label{fAF}%
\end{equation}
Assuming $\varepsilon\propto (1/u)^{1+w}$, we found the implicit solution
\begin{equation}
    \frac{1+w}{2}\kappa_0 t = \frac{\kappa_0}{\kappa}-\mathrm{atanh}\frac{\kappa}{\kappa_0} - \sqrt{2}\arctan\left(\sqrt{2} \frac{\kappa}{\kappa_0}\right)
\end{equation}
for $\kappa(t)$. The expanding branch $\kappa>0$ describes a smooth transition from an expanding de Sitter universe to an expanding Friedmann universe with $a\propto t^{2/3(1+w)}$. Its conformal diagram is given by the upper triangle of the left diagram of figure (\ref{fig:CDfriedmann}), cf. \cite{Mbook}. 
Comoving geodesics start from $\widetilde{i^-}$ and reach $i^+$ after infinite proper time. Other causal geodesics, however, are past incomplete in the same way as in an expanding de Sitter space in the flat slicing. At the line $t=-\infty$ all curvature invariants are bounded and hence we can complete the diagram by gluing the contracting solution $\kappa<0$, which is related to the expanding solution simply by time reversal. Albeit $\phi =t$ is obviously discontinuous at the junction, the metrics can be joined smoothly just like the expanding and contracting de Sitter space. In this way we obtain a geodesically complete, non-singular spacetime.

\paragraph{Non-flat universe:}
Let us now extend our analysis to the spatially non-flat case $\varkappa = \pm 1$. In this case the relation $\kappa^2(u)$ provided by (\ref{mF}) does not have to be bijective. In fact, if it were one-to-one, the same arguments as above would apply and we would run into a curvature singularity as $ {^3\!R}(u\to0)\to \infty$.
This can only be avoided by a bounce at some finite $u_{\min}$.

A zero of the relation $\kappa^2(u)$ at $u_{\min}$ can connect the contracting half-plane $\kappa<0$  to the expanding half-plane $\kappa>0$.  If at this point also $\dot{\kappa}$ would vanish, there would be a static fixed point solution at $(u,\kappa) = (u_0,0)$. Thus, by (\ref{kdot}), only a zero of first order of $\kappa^2(u)$ can describe an actual bounce.

For concreteness, consider the case where $\varepsilon=3c/u^{1+w}$ and make the simple power law ansatz
\begin{equation}
    h({^3\!R}) = - 6\left(\frac{\delta}{6}{^3\!R}\right)^{2n} = - 6\left(\frac{\delta}{a^2}\right)^{2n},
\end{equation}
where the prefactors were chosen for later convenience and we restrict to even powers in order to obtain the same high curvature modifications for both $\varkappa=\pm1$. The right hand side of the modified Friedmann equation (\ref{mF}) then becomes proportional to
\begin{equation*}
    \frac{c}{a^{3(1+w)}} - \frac{\varkappa}{a^{2}} - \left(\frac{\delta}{a^{2}}\right)^{2n}.
\end{equation*}
Since the left hand side of (\ref{mF}) should be a one-to-one function of $\kappa^2$ which is linear in the low curvature limit, a zero of this right hand side means that $\kappa=0$ at such a point. In standard GR, setting $\delta =0$, this right hand side has just one non-trivial zero, namely in the case $\varkappa =1$ at 
\begin{equation}
    a_{\textup{max}} = c^{1/(1+3w)},
    \label{umax}
\end{equation}
corresponding to the moment of recollapse of a closed universe. 
In our modification, by the choice of sign of $h$, there is now also another first order zero describing a bounce. At the moment of bounce the linear contribution of ${^3\!R}$ is negligible compared to $h$ and we find
\begin{equation}
    a_{\textup{min}} = \left(\frac{\delta^{2n}}{c}\right)^{1/(4n-3(1+w))}.
    \label{umin}
\end{equation}
Assuming that $a_{\max} \gg a_{\min}$, there is some intermediate region where the energy density $\varepsilon$ dominates both spatial curvature terms. In this region (\ref{mF}) becomes like in the spatially flat case and we expect a stage where $\kappa^2 \sim \kappa_0^2$ is approximately constant at the limiting curvature. A contracting universe will hence undergo a stage of exponential contraction before going through a bounce followed by exponential expansion.

We can estimate the duration of this inflationary stage which is, by time reversal symmetry, equal to the duration of the stage of exponential contraction. Inflation will end at the moment when $\kappa^2$ and hence also the left hand side of (\ref{mF}) drops below the order of $\kappa_0^2$. This will happen around $a=a_f$ where
\begin{equation}
    a_{f}^{3(1+w)} \sim \frac{c}{\kappa_0^2}.
\end{equation}
The value $a_i$ at the beginning of inflation will not be much different from $a_{\textup{min}}$. The number of e-folds expressed through the dimensionless quantities $\tilde{c}$ and $\tilde{\delta}$ defined by
\begin{equation*}
    \tilde{c} = c \,\kappa_0^{3(1+w)-2},\qquad \tilde{\delta} = \delta \,\kappa_0^{2-1/n},
\end{equation*}
can hence be approximated by
\begin{equation}
    N  \sim  \frac{2n}{4n-3(1+w)}\ln \left(\tilde{c}^{\frac{2}{3(1+w)}} \tilde{\delta}^{-1} \right).
    \label{efolds1}
\end{equation}
A necessary number of e-folds can be achieved e.g. by an exponentially small value of $\tilde{\delta}$ or an exponentially large value of $\tilde{c}$. Note that for radiation with $w=1/3$ we have to choose $n>1$ in order to still have a bounce. On the other hand, for $n=1$, a value smaller than, but close to $1/3$,
\begin{equation}
    w=\frac{1}{3}(1 -\epsilon), \quad \epsilon \ll 1,
\end{equation}
 can also explain a large number of e-folds even when all other dimensionless parameters are of the order of unity. In this case
\begin{equation}
    N \sim  \frac{\ln (\tilde{c}/ \tilde{\delta}^{2} )}{\epsilon}.
    \label{efolds2}
\end{equation}
Let us illustrate this in a simple concrete example for a closed universe ($\varkappa =1$) and with the familiar choice (\ref{fAF}) for the function $f$. In this case (\ref{mF}) becomes
\begin{equation}
    \frac{1}{9}\kappa^{2}\left(  \frac{1+2\left(  \kappa^{2}/\kappa_{0}%
^{2}\right)  }{1-\left(  \kappa^{2}/\kappa_{0}^{2}\right)  }\right)
= \frac{1}{a^2} \left[\left(\frac{a_\textup{max}}{a}\right)^{2-\epsilon}\left(1-\left(\frac{a_\textup{min}}{a}\right)^{\epsilon}\right)-1\right],
\label{nfex}
\end{equation}
where $a_{\max}$ and $a_{\min}$ are defined by (\ref{umax}) and (\ref{umin}).
Let us analyze the asymptotics of this equation starting at the moment of recollapse where we set $t=0$. At this point $a\sim a_{\max} \gg a_{\min}$ and $\kappa \ll \kappa_0$. Moreover, the deviation from $w=1/3$ is irrelevant for the behaviour in this region and we set $\epsilon=0$ for simplicity.  Recalling that $\kappa = 3\dot{a}/a$, (\ref{nfex}) in this case becomes
\begin{equation}
    \dot{a}^2 =\left(\frac{a_\textup{max}}{a}\right)^{2}-1
\end{equation}
and has the solution
\begin{equation}
    a(t) = \sqrt{a_\textup{max}^2-t^2}.
    \label{recolsol}
\end{equation}
This would be the full exact solution of (\ref{nfex}) in standard GR, i.e. if we would set $a_{\min}=0$ and $\kappa_0\to\infty$. It describes a closed universe starting from a Big Bang at $t= -a_{\max}$, expanding until $t=0$, then recollapsing until finally reaching a Big Crunch at $t= a_{\max}$. Since for this solution
\begin{equation}
    \kappa(t) = \frac{3t}{t^2-a^2_\textup{max}},
    \label{ksingsol}
\end{equation}
both Big Bang and Big Crunch represent curvature singularities.
In our theory, however, $\kappa$ becomes of order of the limiting curvature at $t -a_{\max}\sim 1/\kappa_0$  and the modification starts to take over.

In the region where the energy density is dominating both spatial derivative terms, (\ref{nfex}) becomes like for a flat Friedmann universe. The exact implicit solution of an equation of this form was obtained in \cite{AFmimetic} and for the case at hand it reads
\begin{equation}
\frac{2}{3}\kappa_{0}(t-a_{\max})=\frac{\kappa_{0}}{\kappa}-\operatorname{atanh}%
\frac{\kappa}{\kappa_{0}}-\sqrt{2}\arctan\left(  \sqrt{2}\frac{\kappa}%
{\kappa_{0}}\right),  \label{impsol}%
\end{equation}
where the constant of integration was fixed such that the $-\kappa\ll\kappa_0$ asymptotic
\begin{equation}
    \kappa = \frac{3}{2(t-a_{\max})}
    \label{crunchasymp}
\end{equation}
of (\ref{impsol}) matches the $t\sim a_{\max}$ asymptotic of (\ref{ksingsol}).
The contracting branch of (\ref{impsol}) describes a smooth transition between the asymptotic (\ref{crunchasymp}) at $t \ll a_{\max}$ and the asymptotically constant solution $\kappa \sim -\kappa_0$ at $t \gg a_{\max}$. Hence, the scale factor does not vanish at $t= a_{\max}$ but after $t \gg a_{\max}$  starts to decrease exponentially as
\begin{equation}
    a \propto \exp \left(-\tfrac{1}{3}\kappa_0 t\right).
\end{equation}
Contrary to the spatially flat case, the exponential contraction now cannot continue until $t\to\infty$ because at some point the spatial curvature dependent potential will start to counteract this contraction and thus prevent an unbounded growth of the spatial curvature. 
Expanded around $a\sim a_{\min}$ (\ref{nfex}) becomes
\begin{equation}
\kappa^{2}/\kappa_0^2\left(  \frac{1+2\left(  \kappa^{2}/\kappa_{0}%
^{2}\right)  }{1-\left(  \kappa^{2}/\kappa_{0}^{2}\right)  }\right)
=  \tilde{\epsilon} \left(\frac{a}{a_{\min}}-1\right),
\end{equation}
where
\begin{equation}
    \tilde{\epsilon} : = \epsilon\,\frac{9}{ \kappa_0^2\,a_{\min}^2} \left(\frac{a_{\max}}{a_{\min}}\right)^{2-\epsilon}.
\end{equation}
Isolating $a$ from this equation on one side and taking the time derivative of the logarithm of this equation, we obtain a separable first order differential equation for $\kappa$ with the implicit solution 
\begin{equation}
    \frac{1}{6}\kappa_0 (t -t_b) = \mathrm{atanh}\frac{\kappa}{\kappa_0} + \tilde{\epsilon}_{-}\arctan\left(\tilde{\epsilon}_{-} \frac{\kappa}{\kappa_0} \right)+\tilde{\epsilon}_{+}\arctan\left(\tilde{\epsilon}_{+} \frac{\kappa}{\kappa_0} \right) ,
    \label{aminas}
\end{equation}
where the constant of integration $t_b$ corresponds the moment of bounce and
\begin{equation*}
    \tilde{\epsilon}_{\pm}^2 : = \frac{4}{\tilde{\epsilon}-1\pm\sqrt{1-10\tilde{\epsilon}+\tilde{\epsilon}^2}}.
\end{equation*}
The $t\ll t_b$ asymptotic of the contracting branch of this solution is $\kappa \sim -\kappa_0$, in agreement with the late time asymptotic of (\ref{impsol}). Note that the sign in front of the $\mathrm{atanh}$ term in (\ref{aminas}) is opposite to that in (\ref{impsol}). This means that $-\kappa$ is now decreasing again until $\kappa =0$ at $t=t_b$. 
Expansion of (\ref{aminas}) around $\kappa \sim 0$ yields
\begin{equation}
    \kappa = \frac{\tilde{\epsilon}}{6 } \kappa_0 (t-t_b).
\end{equation}
Integrating again, we find a smooth bounce described by the leading order solution
\begin{equation}
a= a_{\textup{min}}\left(1+\frac{\tilde{\epsilon}}{12} \kappa_0 \left(t-t_{b}\right)^2\right)^{1/3}.
\end{equation}
At $t=t_b$ we pass from the contracting into the expanding halfplane. By time reversal invariance, the solution in the expanding plane will be just a mirror image of the solution in the contracting plane. 
Hence, after the bounce $\kappa$ will grow until reaching $\kappa_0$ as described by the expanding branch of (\ref{aminas}). It will stay approximately constant for the number of e-folds given by (\ref{efolds2}) followed by a smooth graceful exit described by the expanding branch of (\ref{impsol}). Finally, after $\kappa\ll\kappa_0$, the term linear in spatial curvature will dominate again and cause a recollapse according to (\ref{recolsol}), restarting the whole cycle anew. The solution is hence eternally oscillating clockwise in the phase space $(u,\kappa)$ on the closed trajectory described by (\ref{nfex}).

\clearpage
\subsection{Bianchi Universes}
The general\footnote{Excluding Kantowski-Sachs type models which bear more similarity with the interior of black holes than with cosmological models, cf. section \ref{OnGen}.} form of a homogeneous but not necessarily isotropic spatial metric is given by 
\begin{equation}
    \gamma_{ab} = \gamma_{AB} \,e^{A}_ae^{B}_b,
\end{equation}
where $\gamma_{AB}$ is spatially constant and the frame one-forms $e^{A}_a$ are constant in time and satisfy
\begin{equation}
     \left ( \frac{\partial e^{C}_a}{\partial x^b} -\frac{\partial e^{C}_b}{\partial x^a}\right )e^a_Ae^b_B = \mathcal{C}^C_{AB}.
     \label{structureconst}
\end{equation}
Here $e^a_A$ is the inverse of $e^{A}_a$ and $\mathcal{C}^C_{AB}$ are the structure constants of the group of motion of the three-dimensional homogeneous space under consideration \cite{Landau}, \cite{Henneaux}.
The inverse metric and the metric determinant are  given by
\begin{equation}
    \gamma^{ab} = \gamma^{AB} \,e_{A}^a e_{B}^b, \qquad \sqrt{\gamma} = u v,
\end{equation}
where $\gamma^{AB}$ is the inverse of $\gamma_{AB}$, $v$ is the determinant of $e^{A}_a$ and $u^2$ is the determinant of $\gamma_{AB}$ and hence depends only on time.

The extrinsic curvature of these homogeneous spatial slices is given by
\begin{equation}
    \kappa_{ab} = \frac{1}{2} \dot{\gamma}_{AB} \,e^{A}_ae^{B}_b=:\kappa_{AB} \,e^{A}_ae^{B}_b, \qquad \kappa^a_{b} = \frac{1}{2} \gamma^{AC}\dot{\gamma}_{CB} \,e_{A}^ ae^{B}_b=:\kappa^A_{B} \,e_{A}^ ae^{B}_b
\end{equation}
and we see that
\begin{equation}
    \kappa = \kappa^a_{a} = \kappa^A_{A} = \frac{\dot{u}}{u}.
\end{equation}
The spatial connection coefficients are given by
\begin{equation}
    \lambda^c_{ab} = \frac{\partial e^C_a }{\partial x^b}\,e^c_C - \mathcal{A}^C_{AB}\,e^c_Ce^A_ae^B_b,
\end{equation}
with the spatially constant coefficients
\begin{equation}
    \mathcal{A}^C_{AB} = \frac{1}{2}\left ( \mathcal{C}^C_{AB} -\mathcal{C}^E_{DB}\gamma_{EA}\gamma^{DC}-\mathcal{C}^E_{DA}\gamma_{EB}\gamma^{DC} \right ).
\end{equation}
Note that $\gamma^{AB}\mathcal{C}^C_{AB}=0$ and $\mathcal{A}^C_{DC} = \mathcal{C}^C_{DC}$. 
The mixed components of the $4-$Ricci tensor are given by
\begin{equation}
    R^0_{a} = \left(\kappa^{D}_{C}\mathcal{C}^C_{AD}-\kappa^{D}_{A}\mathcal{C}^{E}_{DE}\right)e^A_a,
\end{equation}
and the spatial Ricci curvature is
\begin{equation}
    {^3\!R_{ab}} = {^3\!R_{AB}} \,e_{a}^A e^{B}_b,\qquad
    ^3\!R_{AB} = \mathcal{A}^C_{DC}\mathcal{A}^D_{AB}-\mathcal{A}^D_{BC}\mathcal{A}^C_{AD}.
\end{equation}
The spatial Bianchi identity reads
\begin{equation}
    {^3\!R^{D}_{C}}\mathcal{C}^C_{AD}-{^3\!R^{D}_{A}}\mathcal{C}^{E}_{DE} =  0.
\end{equation}
Note that ${^3\!R^a_{a}} =  {^3\!R^A_{A} }= {^3\!R}$ is constant in space. Hence we can immediately solve (\ref{Xisol}) to find that 
\begin{equation}
    \Xi \propto \frac{1}{u},
\end{equation}
and the spatial curvature contribution (\ref{Sab}) to the spatial modified Einstein equation is just
\begin{equation}
    S^{a}_{b} = (1-h') {^3\!R^{a}_{b}}.
\end{equation}
The whole modified Einstein equation can be expressed independent of the frame vectors. Let us restrict to the case of isotropic, comoving matter, e.g. dust or mimetic matter, with total energy density $\varepsilon$. Then the temporal equation (\ref{mEE00}) reads
\begin{equation}
        \frac{1}{3}\left(  f-2\kappa f^{\prime}\right)  \kappa^{2}-\Lambda
+\kappa\Lambda^{\prime}=\frac{1}{2}\left(  f+\kappa f^{\prime}\right)
\tilde{\kappa}_{B}^{A}\tilde{\kappa}_{A}^{B}+ \frac{1}{2}\left(h-{^3\!R}\right) + \varepsilon.
\label{Bianchi00}
\end{equation}
The trace subtracted spatial equations (\ref{mEEabsub}) become
\begin{equation}
    -\frac{1}{u}\,\partial_t\left(u\,f \,\tilde{\kappa}^{A}_{B}\right) =  (1-h')\left({^3\!R^{A}_{B}}- \frac{1}{3}{^3\!R}\,\delta^{A}_{B}\right).
    \label{Bianchiab}
\end{equation}
Contracting with $ \tilde{\kappa}^{B}_{A}$ we find the useful equation
\begin{equation}
    \frac{1}{f u^2}\partial_t \left(u^2 f^2 \tilde{\kappa}^{A}_{B} \tilde{\kappa}^{B}_{A} \right) = - 2 \left(1-h'\right) {^3\!R^{A}_{B}}\,\tilde{\kappa}^{B}_{A}.
    \label{Bianchispcontr}
\end{equation}
Finally, the mixed component equation is simply
\begin{equation}
    \Tilde{\kappa}^{D}_{C}\mathcal{C}^C_{AD}-\Tilde{\kappa}^{D}_{A}\mathcal{C}^{E}_{DE} =  0.
    \label{Oa}
\end{equation}
Let us in the following assume that the frame metric $\gamma_{AB}$ is diagonal. This additional assumption is an expression of non-rotating Kasner axes, cf. \cite{Henneaux}.

\subsubsection{Bianchi type I.} This is the case where all structure constants vanish and the spatial slices are hence Euclidean. (\ref{Bianchiab}) is easily integrated and yields
\begin{equation}
    \tilde{\kappa}^{A}_{B} = \frac{\lambda^{A}_{B}}{fu},
    \label{Bianchi1int}
\end{equation}
with constants of integration $\lambda^{A}_{B}$ and (\ref{Bianchi00}) becomes
\begin{equation}
        \frac{1}{3}\left(  f-2\kappa f^{\prime}\right)  \kappa^{2}-\Lambda
+\kappa\Lambda^{\prime}=\frac{f+\kappa f^{\prime}}{2f^2} \frac{\bar{\lambda}^2}{u^2} + \varepsilon,
\label{Bianchi100}
\end{equation}
where $\bar{\lambda}^2:=\lambda^{A}_{B}\lambda^{B}_{A}$. Like in the Friedmann case, this equation defines an integral curve in the phase space spanned by $u$ and $\kappa$. Provided that $(f+\kappa f')/f^2$ is bounded, it will look qualitatively similar to (\ref{mF}) for the Friedmann universe. 
At low curvatures where $\kappa \ll \kappa_0$ and $u\to \infty$, the right hand side will be dominated by any contribution of matter with equation of state $w<1$. The evolution of $u$ is then given by
\begin{equation}
    u \propto t^{2/(1+w)}
\end{equation}
like in a Friedmann universe. Moreover, according to (\ref{Bianchi1int}), the anisotropic extrinsic curvature contribution $\tilde{\kappa}^{A}_{B} \propto u^{-1}$ will decay at $t \to \infty$ faster than $\kappa \propto t^{-1}$. This means that the presence of any kind of isotropic matter with $w<1$ will eventually lead to isotropy of an expanding universe.

On the other hand, approaching $u\to 0$, we see that the term $u^{-2}$ coming from curvature due to anisotropy will dominate any such matter contribution. This is why in order to understand the behaviour close to singularities it is sufficient to study the vacuum case. 
Without modifications, the vacuum solution is given by the Kasner metric, featuring a singularity. In an attempt to remove this anisotropic singularity, we will find that the property of asymptotic freedom is an inevitable condition for any non-singular modification:

Just as for a modified flat Friedmann universe, (\ref{Bianchi100}) establishes a one-to-one relation between $\kappa$ and $u$. By the same argument as in the last section, the only way it can be non-singular is for $\kappa$ to tend to its constant limiting value as $u\to 0$. While in this case $\kappa^2$ as well as $\dot{\kappa}$ are bounded, 
\begin{equation}
    \tilde{\kappa}_{B}^{A}\tilde{\kappa}_{A}^{B} = \frac{\bar{\lambda}^2}{f^2 u^2}
\end{equation}
will become singular, unless that $f(\kappa)$ diverges fast enough as $\kappa \to \kappa_0$. It follows that the only way to avoid a curvature singularity in (\ref{Ricciscalar}) is a fast enough divergence of $f$ at the limiting curvature. Moreover, if $1/(fu) \to 0$, then the anisotropic Kasner solution will even become isotropic close to the limiting curvature. A concrete illustration of this was presented in \cite{AFmimetic}.

\subsubsection{Bianchi type V.}
As an example for an anisotropic, spatially non-flat universe where we can still understand the modified Einstein equation in terms of a two dimensional phase portrait, we consider Bianchi type $\textup{V}$. Here the non-vanishing structure constants are determined by
\begin{equation}
    \mathcal{C}^{\bar{2}}_{\bar{1}\bar{2}}= 1 ,\qquad \mathcal{C}^{\bar{3}}_{\bar{1}\bar{3}}= 1.
\end{equation}
We calculate
\begin{equation}
    \mathcal{A}^{\bar{2}}_{\bar{1}\bar{2}}= \mathcal{A}^{\bar{3}}_{\bar{1}\bar{3}}=1, \qquad \mathcal{A}^{\bar{1}}_{\bar{2}\bar{2}} = - \gamma^{\bar{1}\bar{1}}\gamma_{\bar{2}\bar{2}}, \qquad \mathcal{A}^{\bar{1}}_{\bar{3}\bar{3}} = - \gamma^{\bar{1}\bar{1}}\gamma_{\bar{3}\bar{3}},
\end{equation}
and find that the spatial curvature components in the frame are given by
\begin{equation}
    {^3\!R^{\bar{1}}_{\bar{1}}} = {^3\!R^{\bar{2}}_{\bar{2}} }={^3\!R^{\bar{3}}_{\bar{3}} } = -2\gamma^{\bar{1}\bar{1}}.
\end{equation}
Hence the spatial curvature is still isotropic and (\ref{Bianchiab}) has the first integral
\begin{equation}
     \tilde{\kappa}^{A}_{B} = \frac{\lambda^{A}_{B}}{fu}.
\end{equation}
By the mixed components (\ref{Oa}) it follows that
\begin{equation}
    \tilde{\kappa}^{\bar{1}}_{\bar{1}} =0, \qquad \tilde{\kappa}^{\bar{2}}_{\bar{2}} = - \tilde{\kappa}^{\bar{3}}_{\bar{3}} =: \frac{\tilde{\lambda}}{fu},
\end{equation}
where we can assume without loss of generality that the constant of integration $\tilde{\lambda}\geq 0$. Integrating again yields the frame metric
\begin{equation}
    \gamma_{AB}  = u^{2/3} \,\mathrm{diag}\left(1/\alpha^2,b^2,\alpha^2 b^{-2}\right),
\end{equation}
where $\alpha$ is a constant of integration and $b$ is a function of time. The differential equations (\ref{structureconst}) are solved by the frame vectors
\begin{equation}
    e^{\bar{1}}_a = \alpha \, \delta^1_a, \quad e^{\bar{2}}_a = e^{\alpha x}  \,\delta^2_a, \quad e^{\bar{3}}_a = e^{\alpha x}/\alpha  \,\delta^3_a
\end{equation}
and, fixing overall constant factors, this frame metric corresponds to the spacetime  metric
\begin{equation}
    \mathrm{d}s^2 = \mathrm{d}t^2 - u^{2/3}\left[\mathrm{d}x^2 + e^{2 \alpha x } \left(b^2\mathrm{d}y^2+b^{-2} \mathrm{d}z^2\right)\right].
\end{equation}
Note that the spatial slices are spaces of constant negative curvature. 
The equation (\ref{Bianchi00}) for this class becomes
\begin{equation}
     \frac{1}{3}\left(  f-2\kappa f^{\prime}\right)  \kappa^{2}-\Lambda
+\kappa\Lambda^{\prime}= \frac{f+\kappa f^{\prime}}{f^2}\frac{\tilde{\lambda}^2}{u^2}+ \frac{1}{2}\left(h(-6 \alpha^2 u^{-2/3})+6 \alpha^2 u^{-2/3}\right) + \varepsilon.
     \label{type5-00}
\end{equation}
It is again just an integral curve in the phase space $(u,\kappa)$ which can be treated as for the non-flat Friedmann universe. If $h$ includes a term $\propto ({^3\!R})^n$ with $n\geq3$ and has the right sign, it will describe a bounce in the same way as (\ref{nfex}). 
The solution for $b$ is determined by
\begin{equation}
    \frac{\dot{b}}{b} = \tilde{\kappa}^{\bar{2}}_{\bar{2}} = \frac{\tilde{\lambda}}{fu} \geq 0. 
    \label{BianchiVb}
\end{equation}
Hence $b$ is monotonically increasing and the moment of greatest slope of $b$ is at the bounce, where $u$ assumes its minimum, $\kappa =0$ and $f(\kappa)=1$. 

Long before/long after the bounce, i.e. at $ u\gg u_{\min}$, the linear contribution of spatial curvature $\propto u^{-2/3}$ is dominating both $h$ and $\tilde{\lambda}^2/u^2$. Moreover, in this region $\kappa^2 \ll \kappa_0^2$. In the case of vacuum, the asymptotic solution of (\ref{type5-00}) is hence given by
\begin{equation}
    u(t) = \left(\alpha \left|t\right|\right)^3.
\end{equation}
Integrating (\ref{BianchiVb}) then yields
\begin{equation}
    b \propto \exp \left(\frac{-\tilde{\lambda}}{2\alpha^3 t^2}\right) \xrightarrow[t\to \pm \infty]{ }  b_0^\pm.
\end{equation}
Fixing the constant of integration, we can achieve that $b_0^+ = 1/b_0^-$. Hence, starting at $t\to -\infty$ from a contracting spacetime, after the bounce we obtain the time reversed expanding spacetime where the directions $y$ and $z$ are interchanged. Since $\kappa^a_b\kappa^b_a $ is everywhere bounded and in the early/late time asympotic it holds that $\kappa^a_b\kappa^b_a\propto \kappa^2 \propto 1/ t^2$, the condition for causal completeness from appendix A is satisfied.

\newpage
\subsubsection{Bianchi types II, VI$_{0}$, VII$_{0}$, VIII, IX.}
These five Bianchi types can be treated on a common footing by labeling the non-vanishing structure constants as
\begin{equation}
    \mathcal{C}^{\bar{1}}_{\bar{2}\bar{3}}=\lambda, \qquad \mathcal{C}^{\bar{2}}_{\bar{3}\bar{1}}=\mu, \qquad \mathcal{C}^{\bar{3}}_{\bar{1}\bar{2}}=\nu.
\end{equation}
The individual classes can then be read off from the following table:
\begin{figure}[h]
    \centering
    \begin{tabular}{c|c|c|c|c|c}
   &II & VI$_{0}$ & VII$_{0}$ & VIII & IX  \\
   \hline
  $\lambda$ & 1 & 1 & 1 & 1 & 1 \\
  $\mu$ & 0 & -1 & 1 & 1 & 1 \\
  $\nu$ & 0 & 0 & 0 & -1 & 1 
\end{tabular}
\end{figure}
\\
\noindent
Corresponding solutions for the frame vectors can be found in \cite{Henneaux}. Taking a diagonal metric, the mixed components (\ref{Oa}) are trivially satisfied. Parametrizing the frame metric as 
\begin{equation}
    \gamma_{AB} = u^{2/3}\mathrm{diag}(a^2,b^2,c^2),
\end{equation}
where $abc=1$, it holds that
\begin{equation}
    \tilde{\kappa}^{\bar{1}}_{\bar{1}} = \frac{\dot{a}}{a},\qquad \tilde{\kappa}^{\bar{2}}_{\bar{2}} = \frac{\dot{b}}{b},\qquad \tilde{\kappa}^{\bar{3}}_{\bar{3}} = \frac{\dot{c}}{c}.
\end{equation}
The spatial curvature components are given by
\begin{align}
    ^3\!R^{\bar{1}}_{\bar{1}} &= \frac{1}{2u^{2/3}}\left(\lambda^2 a^4-(\mu b^2-\nu c^2)^2\right),\\
    ^3\!R^{\bar{2}}_{\bar{2}} &= \frac{1}{2u^{2/3}}\left(\mu^2 b^4 -(\nu c^2-\lambda a^2 )^2\right),\\
    ^3\!R^{\bar{3}}_{\bar{3}} &= \frac{1}{2u^{2/3}}\left(\nu^2c^4-(\lambda a^2-\mu b^2)^2\right).
\end{align}
Finally, the traceless spatial modified Einstein equations (\ref{Bianchiab}) become
\begin{align}
    \partial_t\left(uf \frac{\dot{a}}{a}\right) &=  \frac{(h'-1)u^{1/3}}{3}\left( \lambda a^2 \left(2\lambda a^2 - \mu b ^2 - \nu c^2 \right) - \left(\mu b^2-\nu c^2\right)^2\right), \label{bianchispeq11}\\ 
    \partial_t\left(uf \frac{\dot{b}}{b}\right) &=  \frac{(h'-1)u^{1/3}}{3}\left( \mu b^2 \left(2\mu b^2 - \nu c^2 - \lambda a ^2 \right) - \left(\nu c^2-\lambda a^2\right)^2\right), \\ 
    \partial_t\left(u f \frac{\dot{c}}{c}\right) &=  \frac{(h'-1)u^{1/3}}{3}\left( \nu c^2 \left(2\nu c^2 - \lambda a^2 - \mu b ^2 \right) - \left(\lambda a^2-\mu b^2\right)^2\right).
\end{align}
In general, (\ref{Bianchi00}) describes a hypersurface in a six-dimensional phase space parametrized e.g. by $(u,a,b,\kappa, \tilde{\kappa}^{\bar{1}}_{\bar{1}}, \tilde{\kappa}^{\bar{2}}_{\bar{2}})$. 
Clearly, the general analysis of this system becomes intractable analytically. Let us hence restrict to the case where the conformal degree of freedom decouples from the rest. For general $\lambda,\mu,\nu$ we look for special solutions where (\ref{Bianchi00}) contains only $u$ and $\kappa$ and can be decoupled from the other equations. By (\ref{Bianchispcontr}), the condition for this to be possible is
\begin{equation}
    {^3\!R^{A}_{B}}\,\tilde{\kappa}^{B}_{A}=0\quad \Leftrightarrow \quad \tilde{\kappa}^{A}_{B} \tilde{\kappa}^{B}_{A} \propto \frac{1}{u^2 f^2}.
    \label{decouplingcond}
\end{equation}
After a short calculation, we find that for the Bianchi types under consideration
\begin{equation}
    {^3\!R^{A}_{B}}\,\tilde{\kappa}^{B}_{A} = -\frac{1}{2u^{2/3}}\, \partial_t\left(u^{2/3}\, {^3\!R}\right).
\end{equation}
Hence the condition (\ref{decouplingcond}) is equivalent to
\begin{equation}
    {^3\!R} = \frac{-d}{2u^{2/3}},
\end{equation}
where
\begin{equation}
   d:= \lambda^2a^4+\mu^2b^4+\nu^2c^4-2\left(\mu\nu b^2c^2+\nu\lambda a^2c^2+\lambda \mu a^2b^2 \right ) \overset{!}{=} const.
\end{equation}
The temporal modified Einstein equation is in this case just the same as (\ref{type5-00}) for Bianchi type $\mathrm{V}$. Again, a bounce can be implemented by a term $\propto ({^3\!R})^n$ with $n\geq3$ in $h$. Such a bounce ensures that $\kappa^{A}_{B}\kappa^{B}_{A}$ is bounded. 

In the case of negative spatial curvature ($d>0$) there is only one bounce and no recollapse and hence
\begin{equation}
    \kappa^{A}_{B}\kappa^{B}_{A} \propto \frac{1}{t^2} \qquad \textup{as} \quad t\to \pm \infty,
\end{equation}
like in Bianchi type $\mathrm{V}$. It follows that in the case $d>0$ such a bounce is already enough to ensure causal geodesic completeness, as one finds by slightly modifying the theorem presented in \cite{ChoquetBruhat}, cf. appendix A.

In the case of positive spatial curvature ($d<0$) the solution for $u(t)$ and $\kappa(t)$ will be cyclic, similar to the closed Friedmann universe.
Note that
\begin{equation}
    d+\frac{4\mu\nu}{a^2} = \left(\lambda a^2-\mu b^2 - \nu c^2\right)^2,
\end{equation}
which holds also for simultaneous cyclic permutations of $(a,b,c)$ and $(\lambda,\mu,\nu)$. This can be used to express the right hand side of (\ref{bianchispeq11}) solely through $a$ and $u$. For Bianchi type $\mathrm{IX}$ in particular we find that
\begin{equation}
    \partial_t\left(uf \frac{\dot{a}}{a}\right) =  (1-h')u^{1/3}\left(a^2 \sqrt{d+\frac{4}{a^2}} + \frac{d}{3} \right) .
    \label{typeIXaeq}
\end{equation}
By symmetry, the same equation must also hold if $a$ is replaced with $b$ or $c$. 
Since both $u$ and $\kappa$, which appear as sources in this equation, are periodic in time, we expect that also the solutions for $a$, $b$ and $c$ are oscillating between their minimal and maximal values and the corresponding spacetimes will be non-singular.

\clearpage

\section{Modified Black Hole}
\subsection{Black hole in synchronous coordinates}
In GR the metric of a non-rotating, eternal black hole in the synchronous Lema\^itre coordinates \cite{Lemaitre} is given by
\begin{equation}
    \mathrm{d}s^2 = \mathrm{d}T^2 - \left ( x/x_+ \right )^{-2/3} \mathrm{d}R^2 - \left ( x/x_+ \right )^{4/3} r_g^2\, \mathrm{d}\Omega^2,
    \label{Schwmet}
\end{equation}
where $x = R-T$, and $r_g = 2M$.
These coordinates are regular at the horizon
\begin{equation*}
    x = x_+:= \frac{4}{3} M ,
\end{equation*}
and the region $x>0$ covers both interior and exterior of the Schwarzschild black hole. For comoving observers with $R,\vartheta,\varphi = const.$, $T$ represents proper time. In the Schwarzschild radial coordinate $r=r_g \left(x/x_+\right)^{2/3}$ the paths followed by these synchronous observers correspond to radially infalling geodesics. They start from rest at $r \to \infty$ at proper time $T \to -\infty$ and reach the singularity at $r=0$ at the finite proper time $T=R$.

To see how (\ref{Schwmet}) is modified in our theory, we consider in the synchronous coordinates (\ref{synch}) provided by $T = \phi$ the ansatz
\begin{equation}
    \mathrm{d}s^2 = \mathrm{d}T^2 - a^2\left (x \right ) \mathrm{d}R^2 - b^2\left (x \right ) \mathrm{d}\Omega^2,
    \label{mansatz}
\end{equation}
where the functions $a$ and $b$ still depend only on $x=R-T$. The transformation to Schwarzschild coordinates $t$ and $r$ is given by
\begin{equation}
   t = T-\int\!\mathrm{d}x\,\frac{a^2}{1-a^2}, \qquad r = b(R-T),
   \label{Schwcoord}
\end{equation}
which brings the metric to the form
\begin{equation}
    \mathrm{d}s^2 = (1-a^2) \mathrm{d}t^2 -\frac{a^2}{b'\,^2 (1-a^2)}\mathrm{d}r^2 - r^2\mathrm{d}\Omega^2.
\end{equation}
The dependence of $a$ and $b'$ on $r$ has to be found by inverting $r = b(x)$.
The spatial metric determinant of (\ref{mansatz}) is
\begin{equation*}
    \gamma = a^2b^4\sin^2 \vartheta =: u^2(x) \sin^2 \vartheta,
\end{equation*}
and the non-vanishing components of the extrinsic curvature are given by
\begin{equation*}
    \kappa^R_{R} = \frac{\dot{a}}{a} = - \frac{a'}{a},\qquad \kappa^\vartheta_{\vartheta}=\kappa^\varphi_{\varphi} = \frac{\dot{b}}{b}=-\frac{b'}{b},
\end{equation*}
where the prime denotes $x$-derivatives. 

\clearpage
The spatial Ricci curvature components for the class of metrics (\ref{mansatz}) are given by
\begin{align}
    {^3}\!R^{R}_{R}= R^{R}_{T} &=2\left(\gamma^{RR} \left(\kappa^\vartheta_{\vartheta}\right)^2  - \gamma^{\vartheta\vartheta}+  {^3}\!R^\vartheta _{\vartheta } \right) ,
    \label{eq-3Rrr} \\
    {^3}\!R^\vartheta _{\vartheta } = {^3}\!R^\varphi _{\varphi} &=  \frac{1}{2\kappa^\vartheta _{\vartheta}}\, \left(\gamma^{RR} \left(\kappa^\vartheta_{\vartheta}\right)^2 -\gamma^{\vartheta\vartheta}\right) '- 2\left(\gamma^{RR} \left(\kappa^\vartheta_{\vartheta}\right)^2  - \gamma^{\vartheta\vartheta}\right).  
    \label{eq-3Rthth}
\end{align}
The condition for spatial flatness hence amounts to the single equation
\begin{equation}
    \gamma^{RR} \left(\kappa^\vartheta_{\vartheta}\right)^2  - \gamma^{\vartheta\vartheta}= 0 \quad \quad \Leftrightarrow \quad  a^2 = b'\,^2.
    \label{spflcond}
\end{equation}
In this case the metric in Schwarschild coordinates takes the form
\begin{equation}
    \mathrm{d}s^2 = (1-a^2) \mathrm{d}t^2 -\frac{\mathrm{d}r^2}{(1-a^2)} - r^2\mathrm{d}\Omega^2,
    \label{Schwspfl}
\end{equation}
and we see that the Schwarzschild metric (\ref{Schwmet}) is spatially flat in Lema\^itre coordinates. 

Note that in the direction of the vector field
\begin{equation}
    k^\mu \frac{\partial }{\partial x^\mu} := \frac{\partial }{\partial R} +\frac{\partial }{\partial T} = \frac{\partial }{\partial t} 
    \label{eq-SchwKilling}
\end{equation}
the Lie derivative of (\ref{mansatz}) vanishes. In other words, $k^\mu$ is a Killing vector field with norm
\begin{equation*}
    k^\mu k_{\mu} = 1-a^2(x).
\end{equation*}
It follows that a Killing horizon occurs wherever $a^2(x) = 1$. Let us, in analogy with (\ref{Schwmet}), denote the largest value of $x$ where this happens, i.e. the most exterior horizon, by $x_+$.
We can also calculate the surface gravity $g_s$ of this Killing horizon which is defined by the equation \cite{Poisson}
\begin{equation}
    k^\nu_{\:;\mu} k^{\mu} = g_s \, k^\nu,
\end{equation}
evaluated at the horizon. We find that it is related to the extrinsic curvature of the synchronous slices by
\begin{equation}
    g_s = \kappa^R_{R}(x_+) = -a'(x_+).
    \label{sg}
\end{equation}
\subsection{On generality of the solution $\phi = T$}
\label{OnGen}
In the last section we were making ansatz (\ref{mansatz}) from the beginning in the synchronous coordinates provided by $T = \phi$. One could, however, ask the question how some other synchronous time coordinate would be related to this specific one, i.e. if there is a more general solution of the constraint (\ref{constraint}) in a metric given by (\ref{mansatz}). Such a solution still has to be consistent with the isometries of (\ref{mansatz}), e.g. it should be independent of the angular coordinates by spherical symmetry. Moreover, applying the Lie-derivative $\mathcal{L}_{k^{\mu}\frac{\partial }{\partial x^\mu}}$ to the constraint equation (\ref{constraint}) or to the modified Einstein equation (\ref{mEE}), we find the consistency condition
\begin{equation}
    \left[ k^{\mu}\frac{\partial }{\partial x^\mu}\, ,\, \phi^{,\nu}\frac{\partial }{\partial x^\nu} \right ] =\left (k^\mu\frac{\partial }{\partial x^\mu} \phi^{,\nu} -\phi^{,\mu}\frac{\partial }{\partial x^\mu} k^{\nu}   \right )\frac{\partial }{\partial x^\nu}  =  0,
\end{equation}
from which it follows that
\begin{equation}
    \phi = c\, T + \xi(R-T),
    \label{altsol}
\end{equation}
where $c$ is a constant and $\xi$ is an arbitrary function.
Reinserting into the constraint equation (\ref{constraint}), we find the family of solutions
\begin{equation}
    \xi_\pm^{(c)}(x) = \int\!\mathrm{d}x \frac{c\,a^2 \pm a\sqrt{c^2-1+a^2}}{a^2-1}.
\end{equation}
Introducing the radial coordinate 
\begin{equation}
    \tilde{r} := c\,R - \varrho(R-T)
\end{equation}
and requiring the coordinates $(\tilde{t} :=\phi,\tilde{r},\vartheta,\varphi)$ to be synchronous, i.e. $\tilde{g}^{\tilde{t}\tilde{r}} = 0$, yields the corresponding solution
\begin{equation}
    \varrho_{\pm}^{(c)}(x) = -c\int\!\mathrm{d}x \frac{1\pm c\,a / \sqrt{c^2-1+a^2}}{a^2-1}.
\end{equation}
Note that the combination
\begin{equation}
    \tilde{x} :=\tilde{r}-\tilde{t} = \mp \int\! \mathrm{d}x \frac{a}{\sqrt{c^2-1+a^2}}
\end{equation}
is only a function of $x = R-T$ and hence all functions of $x$ can be expressed as functions of $\tilde{x}$. The full metric ansatz in these coordinates will thus be again of the form
\begin{equation}
    \mathrm{d}s^2 = \mathrm{d}\tilde{t}\,^2 - \tilde{a}^2\left (\tilde{x} \right ) \mathrm{d}\tilde{r}^2 - \tilde{b}^2\left (\tilde{x} \right ) \mathrm{d}\Omega^2,
\end{equation}
where now $\phi = \tilde{t}$ and 
\begin{equation}
    \tilde{a}^2(\tilde{x}) = \frac{c^2-1+a^2(x(\tilde{x}))}{c^2}, \qquad \tilde{b}^2(\tilde{x}) = b^2(x(\tilde{x})).
\end{equation}
Hence, the ansatz (\ref{mansatz}) made above (corresponding to the case $c=1$) is fully general. The Killing vector field $\partial/\partial t$ in these coordinates has the expression
\begin{equation}
   \frac{\partial}{\partial t} = \frac{\partial}{\partial R} +\frac{\partial}{\partial T} = c\left(\frac{\partial}{\partial \tilde{r}} +\frac{\partial}{\partial \tilde{t}}\right).
\end{equation}
Suppose that the metric is spatially flat in the original coordinates $T,R$, i.e. $b'(x)^2 = a^2(x)$. Then the relation in the new coordinates is
\begin{equation}
    \left(\frac{\mathrm{d}\tilde{b}}{\mathrm{d}\tilde{x}}\right)^2 = c^2 \tilde{a}^2(\tilde{x}).
\end{equation}
It follows that there can only be at most one synchronous coordinate system in which such a metric is spatially flat.

The case $c=0$, where even $k^\mu\phi_{,\mu}=0$, is exceptional and deserves special attention. In this case the corresponding synchronous coordinates $(\tilde{t}=\phi, \tilde{r})$ are defined by
\begin{equation}
    \tilde{t} = - \int\!\mathrm{d}x \frac{a}{\sqrt{a^2-1}}, \qquad \tilde{r} = R+\int\!\mathrm{d}x \frac{1}{a^2-1},
\end{equation}
and the metric is
\begin{equation}
    \mathrm{d}s^2 = \mathrm{d}\tilde{t}\,^2 - \left(a^2-1\right) \mathrm{d}\tilde{r}^2 - b^2 \mathrm{d}\Omega^2,
    \label{Kantowski-Sachs}
\end{equation}
where $a$ and $b$ can now be expressed as functions of $\tilde{t}$ only.
This is a Kantowski-Sachs type metric \cite{KantowskiSachs} and hence homogeneous. In these coordinates the metric can never be spatially flat. The condition for spatial flatness in the original $c=1$ coordinates translates to
\begin{equation}
    \left(\frac{\mathrm{d}b}{\mathrm{d}\tilde{t}}\right)^2 =  a^2-1.
\end{equation}
Note that $\phi = \tilde{t}$ is not a global solution of (\ref{constraint}) because it is only defined where $a^2>1$ and becomes singular at the horizon. This is the ansatz that was first considered and then discarded in \cite{BH} before it was realized that the global solution $\phi=T$ has to be used.

\subsection{Modified Einstein equations}
Considering for simplicity the theory where $h=0$, we want to derive the modified Einstein equation for a metric of the form (\ref{mansatz}).
By virtue of the fact that all relevant quantities depend on $R$ and $T$ only through the quantity $x=R-T$, we can replace $\partial_T = - \partial_R$ and reduce partial differential equations to ordinary differential equations. 

For example, for the vacuum case at hand (\ref{Xisol2}) yields
\begin{equation*}
    \Xi = -\tfrac{1}{\sqrt{\gamma}} \int \!\textup{d}T \, \partial_R  \left(  \sqrt{\gamma}R^{R}_{T} \right) =  R^{R}_{T}  = {^3\!R^{R}_{R}},
\end{equation*}
where we set the constant of integration to zero to be consistent with the asymptotic exterior vacuum solution (\ref{Schwmet}).
Hence the temporal modified Einstein equation (\ref{mEE00}) becomes
\begin{equation}
    \frac{1}{3}\left(  f-2\kappa f^{\prime}\right)  \kappa^{2}-\Lambda
+\kappa\Lambda^{\prime}-\frac{1}{2}\left(  f+\kappa f^{\prime}\right)
\tilde{\kappa}_{b}^{a}\tilde{\kappa}_{a}^{b}=\gamma^{RR} \left(\kappa^\vartheta_{\vartheta}\right)^2 - \gamma^{\vartheta\vartheta}.
\label{BH00}
\end{equation}
The trace subtracted spatial equations (\ref{mEEabsub}) read
\begin{equation}
   \frac{1}{u}\left( u f\tilde{\kappa}^{a}_{b}\right)'= {^3\!R^{a}_{b}} - \frac{1}{3}{^3\!R} \delta^{a}_{b}.
   \label{BHspeq}
\end{equation}
By spherical symmetry and tracelessness they contribute only one independent equation. Subtracting the $\vartheta-\vartheta$ equation from the $R-R$ equation and inserting (\ref{eq-3Rrr}), (\ref{eq-3Rthth}) it can be written as
\begin{equation}
   \frac{1}{u} \left(u \,f \left(\frac{b'}{b}-\frac{a'}{a}\right)\right)' = \frac{1}{2\kappa^\vartheta_\vartheta}\left(\gamma^{RR}(\kappa^\vartheta_\vartheta)^2 - \gamma^{\vartheta\vartheta}\right)'.
   \label{speqsub}
\end{equation}
For the Schwarzschild solution (\ref{Schwmet}) it holds that $\kappa = -1/x$. Hence, for large mass black holes with
\begin{equation}
    M \gg \frac{1}{\kappa_0}
\end{equation}
the extrinsic curvature at the horizon $x=x_+ \approx 4M/3$ is much lower than the limiting curvature scale $\kappa_0$ and we can still expect the exterior solution to be given by (\ref{Schwmet}) and modifications to restrict themselves to the interior region. As we have seen, the Schwarzschild solution (\ref{Schwmet}) is exactly spatially flat in the given slicing. Let us assume that the spatial curvature will remain negligible also for some range of $x$ after the modification has taken over. In fact, we will find that the linear contribution of spatial curvature is irrelevant for the region close to the horizon even in the case $M \sim \kappa_0^{-1}$.

In this spatial flatness approximation (\ref{BHspeq}) is easily integrated and yields
\begin{equation}
    \tilde{\kappa}^{R}_{R}=\frac{2M}{fu}, \qquad \tilde{\kappa}^{\vartheta}_{\vartheta}=-\frac{M}{fu},
    \label{excsol}
\end{equation}
where the constants of integration have been fixed to match the Schwarzschild solution in the limit $x\to \infty$. Accordingly, (\ref{BH00}) becomes
\begin{equation}
\frac{\kappa^2\left(f-2\kappa f'\right)-3\left(\Lambda - \kappa \Lambda'\right)}{f+\kappa f'}= \left(\frac{3M}{fu}\right)^2,
\label{BH00sf}
\end{equation}
which is formally the same equation as (\ref{Bianchi100}) for a modified Kasner universe and can be used to determine $u(x)$. We can integrate (\ref{excsol}) again to obtain the solutions for $a(x)$ and $b(x)$ as
\begin{equation}
    a = u^{1/3}\,\left(\frac{2}{3}\,\kappa_0\, e^H\right)^{2/3}, \qquad \qquad b = u^{1/3}\,\left(\frac{2}{3}\,\kappa_0 \, e^H\right)^{-1/3},
    \label{eq-Schwab2}
\end{equation}
where the prefactors have been chosen for dimensionality and later convenience and
\begin{equation}
    H:=\int \! \textup{d}T \, \frac{3M}{f u} .
    \label{H}
\end{equation}
Note that the integrand can be expressed entirely through $\kappa$ from (\ref{BH00sf}). Moreover, applying the same technique as before and taking the time derivative of the logarithm of (\ref{BH00sf}) yields a first order differential equation for $\kappa$ where $M$ drops out. The dependence of $a$, $b$ and $u$ on the mass parameter $M$ can hence only come from a constant of integration which needs to be fixed to match (\ref{Schwmet}). Solutions of the spatially flat approximation hence generically scale as
\begin{equation}
    a,b \propto M^{1/3}.
\end{equation}

For the analogous case of a contracting Kasner universe we have seen in \cite{AFmimetic} that a fast enough divergence of $f$ at the limiting curvature can make anisotropies disappear during contraction. Under the condition that $\dot{H} \propto 1/fu \to 0 $ while $fu^2$ remains finite when the limiting curvature is approached, also for the case at hand it follows that $\tilde{\kappa}^{R}_{R}$ and $\tilde{\kappa}^{\vartheta}_{\vartheta}$ will vanish as $x\to -\infty$. Hence, the functions $a$ and $b$ become alike, up to some finite constant factor
\begin{equation}
    \zeta:= \lim_{x\to-\infty} \frac{a(x)}{b(x)} .
\end{equation}
In the original Schwarzschild solution the function $a$ is increasing as $a \propto b^{-1/2}$ as we go towards $r=b\to0$. Since the modification smoothly connects this solution to $a\propto b$,  it is clear that $a'$ has to change sign at some point where $a$ reaches a maximum value before starting to decrease as we go deeper inside the black hole to $x\to -\infty$. If this maximum value of $a$ is greater than one, we hence expect two Killing horizons $x_\pm$, one on each side of the maximum.
In the limiting case where the maximum is exactly equal to one, these two horizons merge and the region where the Killing vector $k^\mu$ is spacelike (region $\textup{II}$ in the conformal diagram, cf. section \ref{sec:CD}) shrinks to a single horizon. Decreasing the mass of the black hole even further, we find that no horizon occurs at all. Hence there is a minimal mass of order
\begin{equation}
    M_{\min} \sim \frac{1}{\kappa_0}
\end{equation}
below which no black hole solution exists. Moreover, by (\ref{sg}) it follows that the surface gravity of this minimal black hole vanishes. This indicates that the final product of black hole evaporation will approach a minimal mass remnant for which Hawking radiation stops \cite{BlackHoleRemnants}, \cite{Mbook2}.

Before illustrating this in a concrete example, let us discuss the fate of the singularity of (\ref{Schwmet}) at the ``center'' $x=0$.
The asymptotic solution of the spatially flat approximation in the limit $x\to-\infty$  becomes
\begin{equation}
    \mathrm{d}s^2 = \mathrm{d}T^2 - u^{\frac{2}{3}}(x) \left (\zeta^{\frac{4}{3}}\mathrm{d}R^2 +\zeta^{-\frac{2}{3}} \mathrm{d}\Omega^2  \right ),
    \label{asympsol}
\end{equation}
where
\begin{equation}
    u =u_0\exp{\left(\kappa_0 x\right)}.
\end{equation}
The spatial curvature components of this asymptotic solution are given by
\begin{equation}
    {^3}\!R^{R}_{R}= 0,\qquad
    {^3}\!R^\vartheta _{\vartheta } = {^3}\!R^\varphi _{\varphi} =  \left ( 1-\left ( \frac{\kappa_0}{3\zeta} \right )^2 \right ) \left ( \frac{\zeta}{u} \right )^{2/3} .
\end{equation}
Note that for modifications where it happens that $\zeta=\kappa_0/3$, the prefactor in the spatial curvature exactly cancels and the asymptotic solution is hence spatially flat. As a consequence, the full Ricci scalar of this asymptotic solution has a constant value and it describes a part of de Sitter spacetime. This can be seen also by defining the new radial coordinate $\Tilde{R} : =  (u_0 e^{\kappa_0 R} /\zeta)^{1/3} $  which brings (\ref{asympsol}) to the form
\begin{equation}
    \mathrm{d}s^2 = \mathrm{d}T^2 - e^{-\frac{2}{3}\kappa_0 T} \left (\left(\frac{3\zeta}{\kappa_0}\right)^2\mathrm{d}\tilde{R}^2 +\tilde{R}^2 \mathrm{d}\Omega^2  \right ).
\end{equation}
Since inside the inner horizon $x_-$ the Killing vector field $k^\mu$ is timelike again, we know that the metric in this region of the modified solution has to be static. Transforming to static Schwarzschild coordinates (\ref{Schwcoord}), we find the relations $a = \zeta r$, $b' = (\kappa_0/3)  r$ and the asymptotic metric
\begin{equation}
    \mathrm{d}s^2 = (1-\zeta^2r^2) \mathrm{d}t^2 -\frac{(3\zeta/\kappa_0)^2}{(1-\zeta^2r^2)}\mathrm{d}r^2 - r^2\mathrm{d}\Omega^2.
\end{equation}
In the case $\zeta = \kappa_0/3$, the solution in the region $x\in(-\infty,x_-)$ (region $\textup{IIa}$ in figures \ref{fig:CDs} and \ref{fig:CDmax}) hence approaches the static patch of the de Sitter spacetime and has the same causal structure. The singularity is hence replaced by a smooth transition to a part of de Sitter space \cite{ThroughBH}. 

The above shows that in principle it is possible to find a modification for which the spatial curvature and hence also the potential $h({^3\!R})$ will never become important and even exactly vanish in both limits $x\to\pm\infty$.

If $\zeta \neq \kappa_0/3$, the spatial curvature in the asymptotic region is of order
\begin{equation}
    \gamma^{RR} \left(\tilde{\kappa}^{\vartheta}_{\vartheta}\right)^2 -\gamma^{\vartheta\vartheta}\to \frac{\kappa_0/3\zeta-1}{b^2} \propto \frac{1}{u^{2/3}}.
    \label{eq-asympspcont}
\end{equation}
Contracting (\ref{BHspeq}) with $\kappa^b_a$ we find that
\begin{equation}
    \frac{1}{2u^2f}\left(u^2 f^2 \tilde{\kappa}^a_b\tilde{\kappa}^b_a\right)' = \kappa^a_b \,{^3\!\widetilde{R}^b_a} = \frac{1}{3}\left(\frac{\kappa^R_R}{\kappa^{\vartheta}_{\vartheta}}-1\right) \left(\gamma^{RR} \left(\kappa^{\vartheta}_{\vartheta}\right)^2 -\gamma^{\vartheta\vartheta}\right)'.
\end{equation}
By asymptotic freedom and isotropy of the asymptotic solution, the right hand side remains negligible for the whole range of $x$ even if $\zeta\neq \kappa_0/3$. As a consequence it still holds approximately that
\begin{equation}
    \tilde{\kappa}^a_b\tilde{\kappa}^b_a \propto \frac{1}{u^2 f^2},
\end{equation}
and thus the linear spatial curvature contribution $\propto u^{-2/3}$ to (\ref{BH00sf}) is dominated by $u^{-2}$. Hence, we expect the solution in this case to remain qualitatively unchanged compared to the spatially flat approximation except for the fact that now $\left|{^3\!R}\right| \to \infty$ as $x\to-\infty$. 
Naively treating (\ref{BH00}) in the region where $a$ and $b$ are alike as a formal analogue of a modified  non-flat Friedmann universe, one would come to the conclusion that in order to achieve a bounce and prevent this blowing up of spatial curvature, it would take a spatial curvature dependent potential including a term
\begin{equation}
    h = -\left|{^3\!R}\right|^n, \qquad n > 3.
\end{equation}
Of course this argument is purely heuristic and a rigorous verification would require an analysis of the full system of equations given by the analogues of (\ref{BH00}) and (\ref{BHspeq}) with $h\neq 0$. Without simplifying approximations, these constitute a highly coupled system of non-linear differential equations for $a$ and $b$, or alternatively, $u$ and $H$. To thoroughly verify our above speculation would hence require further investigation (perhaps numerical) beyond the scope of this paper. 

\subsection{A spatially flat exact solution}
\label{sec:Solution}
If we can find a modification such that the solution of the modified Einstein equation in the spatial flatness approximation everywhere exactly satisfies the spatial flatness condition (\ref{spflcond}), this would be an exact solution of the full modified Einstein equation, even in the case $h\neq0$. The following modification provides a concrete (perhaps not the simplest) example where this possibility is realized.

Consider the asymptotically free modification given by
\begin{align}
    f(\kappa) &= \frac{1+3\left (\kappa/\kappa_0 \right )^2}{\left ( 1+\left ( \kappa/\kappa_0 \right )^2 \right )\left ( 1-\left ( \kappa/\kappa_0  \right )^2 \right )^2}, \\
    \Lambda(\kappa) &= \kappa^2\left ( \frac{\tfrac{4}{3}\left ( \kappa/\kappa_0  \right )^2}{\left ( 1-\left ( \kappa/\kappa_0  \right )^2\right )^2} \,-\,\frac{1+2\left ( \kappa/\kappa_0  \right )^2}{1+4\left ( \kappa/\kappa_0  \right )^2+3\left ( \kappa/\kappa_0  \right )^4}\right ) + \nonumber\\ & \qquad -\frac{\kappa_0}{6} \kappa\left ( \arctan \frac{\kappa}{\kappa_0} -3\sqrt{3}\arctan\left (\sqrt{3} \, \frac{\kappa}{\kappa_0}  \right )+2\,\mathrm{atanh} \frac{\kappa}{\kappa_0}  \right ).
\end{align}
With this choice (\ref{BH00sf}) becomes
\begin{equation}
    \frac{\kappa^2}{\left(1-\left(\kappa/\kappa_0\right)^4\right)^2}= \left(\frac{3M}{u}\right)^2 .
\end{equation}
Taking the time derivative of the logarithm of this equation we find that
\begin{equation}
    \dot{\kappa} = -\kappa^2\frac{1-\left(\kappa/\kappa_0\right)^4}{1+3\left(\kappa/\kappa_0\right)^4},
    \label{eq-kdot}
\end{equation}
which has the implicit solution
\begin{equation}
    -\kappa_0 \,x =  \frac{\kappa_0}{\kappa} - 2\,\mathrm{atanh}\frac{\kappa}{\kappa_0}  +2\arctan\frac{\kappa}{\kappa_0}.
\end{equation}
Evaluating (\ref{H}) as an integral over $\kappa$ yields
\begin{equation}
    H(\kappa) = \ln \left(-\left(\kappa/\kappa_0\right) \;\frac{1+\left(\kappa/\kappa_0\right)^2}{1+3\left(\kappa/\kappa_0\right)^2}\right),
\end{equation}
where the constant of integration was fixed to match the Schwarzschild solution.
It follows that
\begin{equation}
   \frac{a}{b} = \frac{2}{3} \kappa_0 e^H = \frac{2}{3} \left(-\kappa \;\frac{1+\left(\kappa/\kappa_0\right)^2}{1+3\left(\kappa/\kappa_0\right)^2}\right) = -\frac{1}{3}\left(\kappa -\frac{3M}{fu} \right) = \left|\kappa^{\vartheta}_{\vartheta}\right|,
\end{equation}
which shows that this solution is spatially flat and hence an exact solution of the full modified Einstein equation.

The solutions (\ref{eq-Schwab2}) for $a$ and $b$ expressed through $\kappa$ are given by
\begin{align}
    a^3(\kappa) &= \frac{4M}{3} \, \left|\kappa\right| \,\left(1-\left(\kappa/\kappa_0\right)^4\right)\, \left(\frac{1+\left(\kappa/\kappa_0\right)^2}{1+3\left(\kappa/\kappa_0\right)^2}\right)^2,\\
    b^3(\kappa) &= \frac{9M}{2\kappa^2}\, \left(1-\left(\kappa/\kappa_0\right)^2\right)\left(1+3\left(\kappa/\kappa_0\right)^2\right).
\end{align}
For this particular solution $a$ assumes its maximum value at $\kappa = \kappa_*= -\kappa_0/\sqrt{5}$. At this point
\begin{equation}
    a(\kappa_*) = \left(\frac{18 \kappa_0 }{25 \sqrt{5}} M\right)^{1/3} =:  \left(\frac{M}{M_{\min}} \right)^{1/3},
\end{equation}
and we find that the minimal possible black hole mass in this specific modification is given by
\begin{equation}
    M_{\min} =\frac{25 \sqrt{5}}{18\kappa_0}.
\end{equation}

This solution was studied already in \cite{BlackHoleRemnants}. 
To aid our intuition, let us transform to Schwarzschild coordinates (\ref{Schwcoord}). By virtue of spatial flatness, it takes there the form (\ref{Schwspfl}). The location of the maximum of $a$ in the Schwarzschild $r$-coordinate is given by
\begin{equation}
    r_{\ast }= b(\kappa_*)=\left( 144M/5\kappa _{0}^{2}\right)^{1/3}.
\end{equation}
\clearpage
Far away from the black hole, in the limit $r\rightarrow \infty$, where $(\kappa / \kappa_0)^{2}\ll 1$, we find the expansion 
\begin{equation}
1-a^{2}=1-\frac{2M}{r}\left[ 1-\frac{5}{16}\left( \frac{r_{\ast }}{r}\right)
^{3}+\mathcal{O}\left( \left( \frac{r_{\ast }}{r}\right) ^{6}\right) \right].
\label{19}
\end{equation}%
It follow that the location of the outer horizon of a large mass black hole is given by
\begin{equation}
r_{+}=2M\left[ 1-\frac{729}{6250}\left( \frac{M_{\min }}{M}\right)
^{2}+\mathcal{O}\left( \left( \frac{M_{\min }}{M}\right) ^{4}\right) \right].
\label{20}
\end{equation}%

On the other hand, close to the limiting curvature $\kappa^2\rightarrow \kappa_0^{2}$ we find the expansion
\begin{equation}
1-a^{2}=1-(\zeta r)^{2}\left[ 1-\frac{4}{5}\left( \frac{r}{r_{\ast }}\right)
^{3}+\mathcal{O}\left( \left( \frac{r}{r_{\ast }}\right) ^{6}\right) \right],
\label{21}
\end{equation}%
where $\zeta =\kappa _{0}/3$ and the inner horizon ($ \sim$ de Sitter horizon) occurs at 
\begin{equation}
r_{-}=\zeta^{-1}\left[ 1+\frac{27\sqrt{5}}{1600}\frac{M_{\min }}{M}+\mathcal{O}\left(
\left( \frac{M_{\min }}{M}\right) ^{2}\right) \right] .  \label{22}
\end{equation}%
Both asymptotics fail to describe the region between the two horizons. Expanding the solution around the maximum of $a$ at $r_*$ we find that
\begin{equation}
    1-a^2 \approx 1-\left(\frac{M}{M_{\min}}\right)^{2/3}\left(1-\frac{10}{7}\left(1-r/r_*\right)^2\right).
\end{equation}
For the minimal Black Hole $M=M_{\min}$ inner and outer horizon coincide, i.e. $r_*= r_+ = r_-$, and the metric close to this single horizon is given by
\begin{equation}
    1-a^2 \approx \frac{10}{7}\left(1-r/r_*\right)^2.
\end{equation}
Note the similarity to the near horizon metric of an extremal Reissner-Nordstr\"om black hole.

\clearpage
\subsection{Conformal diagrams}
\label{sec:CD}
The conformal diagrams of the family of solutions found in the last section can be obtained by standard methods by gluing the diagrams of the individual regions separated by horizons, \cite{Mbook2}. For the case of a non-minimal black hole $M>M_{\min}$ with two separate horizons, the solution with range $x\in(-\infty,\infty)$, i.e. $\kappa \in (-\kappa_0,0)$, covers the three regions of the eternal black hole solution, the exterior $\mathrm{I}$, the region between horizons $\mathrm{II}$ and the region $\mathrm{IIa}$ between the inner horizon and $r=0$ which is essentially a static de Sitter patch. By time reversal invariance of our theory, the corresponding white hole solution can be found simply by reversing the arrow of time. Identifying the black and white hole exterior regions, we find the new regions $\mathrm{IV}$ and $\mathrm{IVa}$ which are just time reversed versions of $\mathrm{II}$ and $\mathrm{IIa}$. Note that the static regions $\mathrm{I}$ and $\mathrm{IIa}$ are identical to their time reversed version.
The conformal diagrams encompassing these regions for the three cases $M>M_{\min}$, $M=M_{\min}$, $M<M_{\min}$ are shown in figure \ref{fig:CDs}.

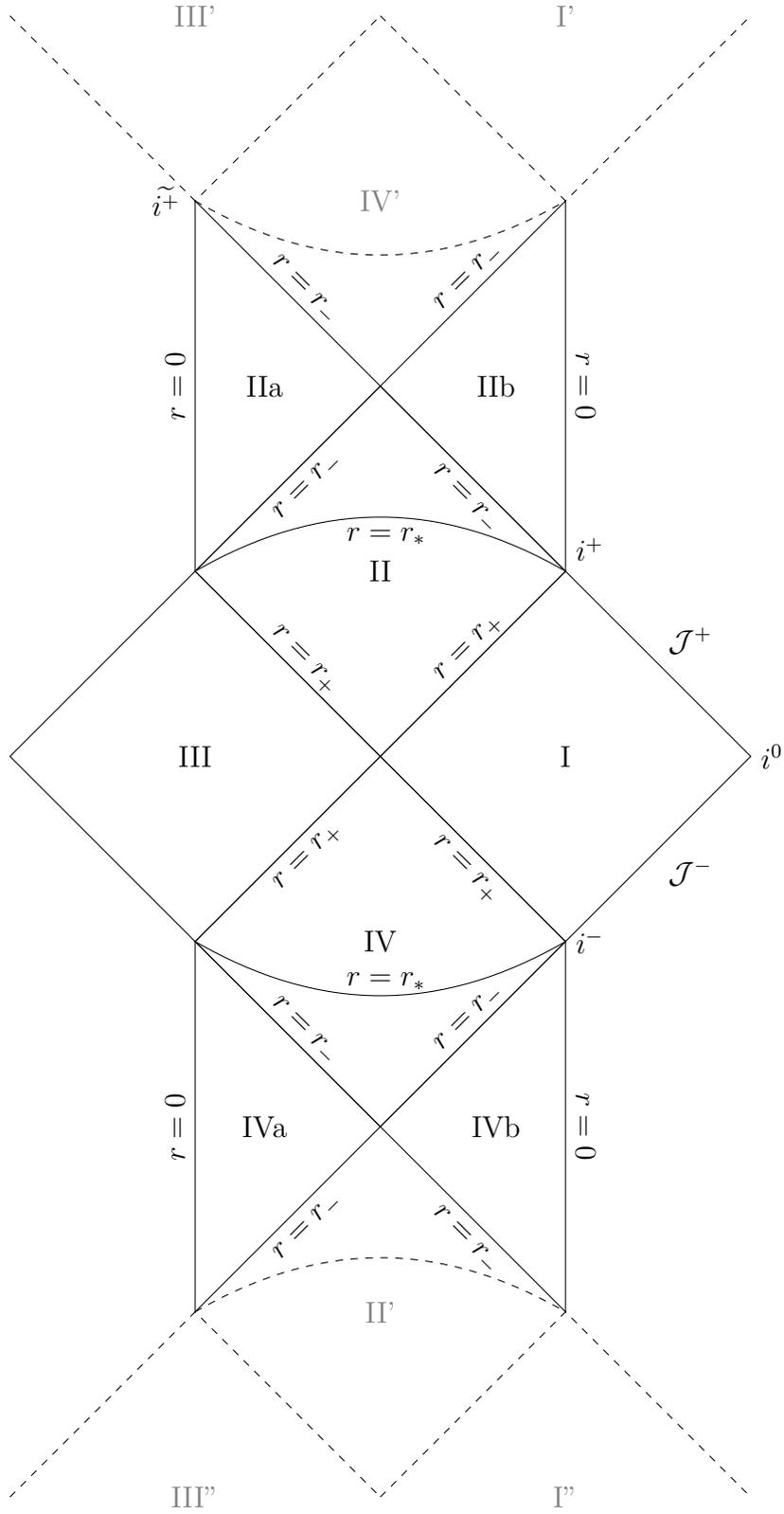
\begin{figure}
\vspace{-4.5cm}
\centering
\begin{tikzpicture}[scale=0.65]
\node (I)    at ( 4,0)   {I};
\node (II)  at (0, 4) {II};
\node (III)   at (-4,0)   {III};
\node (IV)   at (0,-4) {IV};
\node (V)    at (4,8) {};
\node (VI)    at (-4,8) {};
\node (VII)    at (4,-8) {};
\node (VIII)    at (-4,-8) {};
\node (IV')    at (0,12) {};
\node (I')    at (4,16) {};
\node (III')    at (-4,16) {};
\node (II')    at (0,-12) {};
\node (I'')    at (4,-16) {};
\node (III'')    at (-4,-16) {};
\tikzmath{\nd = 11.5mm;} 
       
\path (II) +(2.5,4)  coordinate[label= center:$\textup{IIb}$]+(-2.5,4) coordinate[label=center:$\textup{IIa}$];
       
\path  (IV) +(2.5,-4)  coordinate[label= center:$\textup{IVb}$] +(-2.5,-4) coordinate[label=center:$\textup{IVa}$];

\path (I) +(90:4)  coordinate[label=above right:$i^+$] (Itop)+(-90:4) coordinate[label=right:$i^-$] (Ibot)+(180:4) coordinate (Ileft)+(0:4)   coordinate[label=right:$i^0$] (Iright);
 
\path (II) +(90:4)  coordinate (IItop)+(-90:4) coordinate (IIbot)+(180:4) coordinate (IIleft)+(0:4)   coordinate (IIright);

\path (III) +(90:4)  coordinate(IIItop)+(-90:4) coordinate(IIIbot)+(0:4)   coordinate (IIIright) +(180:4) coordinate (IIIleft);

\path (IV) +(90:4)  coordinate (IVtop)+(-90:4) coordinate (IVbot)+(180:4) coordinate (IVleft)+(0:4)   coordinate (IVright);
       
\path (V) +(90:4)  coordinate (Vtop)+(-90:4) coordinate (Vbot)+(180:4) coordinate (Vleft)+(0:4)   coordinate (Vright);
       
\path (VI) +(90:4)  coordinate[label={[label distance=-7.5mm]0:{$\widetilde{i^+}$}}] (VItop)+(-90:4) coordinate (VIbot)+(180:4) coordinate (VIleft)+(0:4)   coordinate (VIright);
       
\path (VII) +(90:4)  coordinate (VIItop)+(-90:4) coordinate (VIIbot)+(180:4) coordinate (VIIleft)+(0:4)   coordinate (VIIright);

\path (VIII) +(90:4)  coordinate (VIIItop) +(-90:4) coordinate (VIIIbot)+(180:4) coordinate (VIIIleft)+(0:4)   coordinate (VIIIright);

\path (IV')coordinate[label= {[gray]center:$\textup{IV'}$}] +(90:4)  coordinate (IV'top)+(-90:4) coordinate (IV'bot)+(180:4) coordinate (IV'left)+(0:4)   coordinate (IV'right);

\path (I')coordinate[label= {[gray]center:$\textup{I'}$}]  +(90:4)  coordinate (I'top)+(-90:4) coordinate (I'bot)+(180:4) coordinate (I'left)+(0:4)   coordinate (I'right);
\path (III')coordinate[label= {[gray]center:$\textup{III'}$}]  +(90:4)  coordinate (III'top)+(-90:4) coordinate (III'bot)+(180:4) coordinate (III'left)+(0:4)   coordinate (III'right);

\path (II')coordinate[label= {[gray]center:$\textup{II'}$}] +(90:4)  coordinate (II'top)+(-90:4) coordinate (II'bot)+(180:4) coordinate (II'left)+(0:4)   coordinate (II'right);

\path (I'')coordinate[label= {[gray]center:$\textup{I''}$}]  +(90:4)  coordinate (I''top)+(-90:4) coordinate (I''bot)+(180:4) coordinate (I''left)+(0:4)   coordinate (I''right);
\path (III'')coordinate[label= {[gray]center:$\textup{III''}$}]  +(90:4)  coordinate (III''top)+(-90:4) coordinate (III''bot)+(180:4) coordinate (III''left)+(0:4)   coordinate (III''right);

\draw  (Ileft) -- (Itop) -- node[midway, above right]{$\cal{J}^+$}(Iright) -- node[midway, below right]    {$\cal{J}^-$}(Ibot) -- (Ileft) -- cycle;

\draw  (IItop) -- node[midway, below, sloped]{$\phantom{_-}r=r_-$}(IIright) -- node[midway, above, sloped,label={[label distance=-\nd]90:\rotatebox{45}{$\phantom{_+}r=r_+$}}]{ }(IIbot) --node[midway, above, sloped,label={[label distance=-\nd]90:\rotatebox{-45}{$\phantom{_+}r=r_+$}}]{ } (IIleft) --node[midway, below, sloped]{$\phantom{_-}r=r_-$} (IItop) ;

\draw (IIIleft) --  (IIItop) --(IIIright) -- (IIIbot) --(IIIleft) -- cycle;

\draw  (Vbot) -- (Vleft) -- node[midway, above, sloped,label={[label distance=-\nd]90:\rotatebox{45}{$\phantom{_-}r=r_-$}}]{ } (Vtop) -- node[midway, above, sloped] {$r=0$} (Vbot) --cycle;

\draw  (VItop) -- node[midway, above, sloped,label={[label distance=-\nd]90:\rotatebox{-45}{$\phantom{_-}r=r_-$}}]{ }(VIright) -- (VIbot) --node[midway, above,sloped] {$r=0$} (VItop)-- cycle;

\draw (IIleft) to[bend left] node[midway, below,sloped] {$\phantom{_*}r=r_*$} (IIright);
\draw (IVleft) to[bend right] node[midway, above, sloped,label={[label distance=-3.5mm]{$\phantom{_*}r=r_*$}}]{ } (IVright);

\draw  (IVtop) -- node[midway, below, sloped]{$\phantom{_+}r=r_+$}(IVright) -- node[midway, above, sloped,label={[label distance=-\nd]90:\rotatebox{45}{$\phantom{_-}r=r_-$}}]{ }(IVbot) --node[midway, above, sloped,label={[label distance=-\nd]90:\rotatebox{-45}{$\phantom{_-}r=r_-$}}]{ } (IVleft) --node[midway, below, sloped]{$\phantom{_+}r=r_+$} (IVtop) ;

\draw  (VIIbot) -- node[midway, below ,sloped]{$\phantom{_-}r=r_-$}(VIIleft) -- (VIItop) -- node[midway, above, sloped] {$r=0$} (VIIbot) --cycle;

\draw  (VIIItop) -- (VIIIright) -- node[midway, below ,sloped]{$\phantom{_-}r=r_-$}(VIIIbot) --node[midway, above,sloped] {$r=0$} (VIIItop)-- cycle;

\draw[dashed,path fading=north]  (I'left)-- (I'bot) -- (I'right);
\draw[dashed,path fading=north]  (III'left)-- (III'bot) -- (III'right);
\draw[dashed,path fading=south] (IV'left) to[bend right] (IV'right);

\draw[dashed,path fading=south]  (I''left)-- (I''top) -- (I''right);
\draw[dashed,path fading=south]  (III''left)-- (III''top) -- (III''right);
\draw[dashed,path fading=north] (II'left) to[bend left] (II'right);

\end{tikzpicture}
\vspace{-2cm}
\caption{Conformal diagram of the maximally extended black hole solution from section \ref{sec:Solution} in the case $M>M_{\min}$.}
\label{fig:CDmax}
\end{figure}

Note that a static de Sitter patch is not geodesically complete and hence neither are the diagrams $M\geq M_{\min}$ in figure \ref{fig:CDs}. Synchronous observers with $R=const.$ start at $i^-$ ($T=-\infty$) from rest, pass outer and inner horizon and after infinite proper time reach $\widetilde{i^+}$ ($T=\infty, r=0$). Hence these comoving geodesics are complete and fully contained in the union of regions $\mathrm{I}$, $\mathrm{II}$, $\mathrm{IIa}$. However, light rays and massive particles with negative initial radial velocity at $i^-$ will reach $r=0$ at finite synchronous time. Since no singularity occurs at this point, they will simply be reflected towards the upper horizon $r=r_-$ of region $\textup{IIa}$ where also $T=\infty$. The diagram can be easily completed by identifying the black hole region $\textup{IIa}$ with the region $\textup{IVa'}$ of another white hole and continuing this procedure ad infinitum. The conformal diagram of the maximally extended eternal black hole solution is shown in figure \ref{fig:CDmax}.

The maximally extended solution shows that all non-comoving particles falling through the event horizon will eventually escape to another universe \cite{ThroughBH}. It follows that no information is ``trapped'' inside the finite region $\mathrm{IIa}$ and there is no upper limit on the amount of information that can fall into the black hole.

Even though figure \ref{fig:CDmax} bears similarity with the conformal diagram of the Reissner-Nordstr\"om and Kerr spacetimes \cite{FrolovBH}, there is a crucial difference: There is no Cauchy horizon at $r=r_{-}$ and the regions $\mathrm{IIa}$, $\mathrm{IIb}$, $\mathrm{IVa}$, $\mathrm{IVb}$ etc. represent static patches of de Sitter space.  Moreover, there is no singularity at $r=0$ and geodesics reaching this point will simply be reflected.

\clearpage
\begin{figure*}[h]
\centering
\begin{tikzpicture}[scale=0.25]
\tikzmath{\dx = 20;} 

\node (I)    at ( 4,0)   {I};
\node (II)   at (-4,0)   {};
\node (III)  at (0, 4) {II};
\node (IV)   at (0,-4) {IV};
\node (V)    at (4,8) {};
\node (VI)    at (-4,8) {};
\node (VII)    at (4,-8) {};
\node (VIII)    at (-4,-8) {};
\node at (0,-16) {$M>M_{\min}$};

\path (II) +(90:4)  coordinate[label= left:$ $]  (IItop)+(-90:4) coordinate[label=left:$ $] (IIbot)+(0:4) coordinate (IIright)+(180:4) coordinate[label=180:$ $] (IIleft);
\draw[dashed] (IIbot) -- (IIleft) -- (IItop);

\path(I) +(90:4)  coordinate[label=above right:$ $] (Itop)+(-90:4) coordinate[label=right:$ $] (Ibot)+(180:4) coordinate (Ileft)+(0:4)   coordinate[label=right:$ $] (Iright);
       
\path (III) +(2.5,4)  coordinate[label= center:$ $]+(-2.5,4) coordinate[label=center:$\textup{IIa}$];
       
\path(IV) +(2.5,-4)  coordinate[label= center:$ $] +(-2.5,-4) coordinate[label=center:$\textup{IVa}$];
       
\path (III) +(90:4)  coordinate (IIItop)+(-90:4) coordinate (IIIbot)+(180:4) coordinate (IIIleft)+(0:4)   coordinate (IIIright);
       
\path(V) +(90:4)  coordinate (Vtop)+(-90:4) coordinate (Vbot)+(180:4) coordinate (Vleft)+(0:4)   coordinate (Vright);

\path(VI) +(90:4)  coordinate[label=above:$ $]  (VItop)+(-90:4) coordinate (VIbot)+(180:4) coordinate (VIleft)+(0:4)   coordinate (VIright);

\path(VII) +(90:4)  coordinate (VIItop)+(-90:4) coordinate (VIIbot)+(180:4) coordinate (VIIleft)+(0:4)   coordinate (VIIright) ;

\path(VIII) +(90:4)  coordinate (VIIItop)+(-90:4) coordinate (VIIIbot)+(180:4) coordinate (VIIIleft)+(0:4)   coordinate (VIIIright);

\draw  (Ileft) -- (Itop) -- node[midway, above right]    {$\cal{J}^+$}(Iright) -- node[midway, below right]    {$\cal{J}^-$}(Ibot) -- (Ileft) -- cycle;

\draw  (IIItop) -- (IIIright);
\draw (IIIbot) -- (IIIleft);

\draw  (Ileft) -- (VIIItop);
\draw (VIIIright) -- (Ibot);

\draw[dashed]  (Vleft) -- (Vtop) --  (Vbot);

\draw  (VItop) -- node[midway, above right]    { } (VIright) -- (VIbot) --node[midway, above,sloped] {$r=0$} (VItop)-- cycle;

\draw[dashed]  (VIItop) --(VIIbot) -- (VIIleft);

\draw  (VIIItop) -- (VIIIright) -- node[midway, below right]    { }(VIIIbot) --node[midway, above,sloped] {$r=0$} (VIIItop)-- cycle;

\node (0)    at (\dx+1-6,6)    {};
\node (I)    at ( \dx+1,0)   {I};
\node (II)   at (\dx+1-6,-6)   {};
\node (III)  at (\dx+1+6, 6) {};
\path (I) +(-3.75,6)  coordinate[label= center:$\textup{IIa}$]+(-3.75,-6) coordinate[label=center:$\textup{IVa}$];
       
\node at (\dx,-16) {$M=M_{\min}$};

\path (0) +(0:0)  coordinate (00)+(90:6)  coordinate[label= above:$ $] (0top)+(-90:6) coordinate  (0bot)+(90:-6) coordinate (0left)+(0:6)   coordinate (0right) ;
       
\path (II) +(0:0)  coordinate (II0)+(90:6)  coordinate (IItop)+(-90:6) coordinate (IIbot)+(90:-6) coordinate (IIleft)+(0:6)   coordinate (IIright);
       
\path (I) +(90:6)  coordinate (Itop)+(-90:6) coordinate[label= below right:$ $] (Ibot)+(180:6) coordinate (Ileft)+(0:6)   coordinate (Iright);

\draw  (0top) -- node[midway, above right]    { } (0right) -- (0bot) --node[midway, above,sloped] {$r=0$}(0left) -- cycle;

\draw  (IItop) -- (IIright) --node[midway, below right]  { } (IIbot) -- (IIleft) -- cycle;

\draw (Itop) --node[midway, above right]    {$\cal{J}^+$} (Iright) -- node[midway, below right]    {$\cal{J}^-$}(Ibot);

\node (0)    at (2*\dx-6,0)    {};
\node (I)    at (2*\dx-1,0)     {I};
\node at (2*\dx-1,-16) {$M<M_{\min}$};

\path(0) +(0:0)  coordinate (00)+(90:12)  coordinate[label= above:$ $] (0top)+(-90:12) coordinate  (0bot)+(0:12)   coordinate (0right);

\draw  (0top) -- node[midway, above right]    {$\cal{J}^+$} (0right) -- node[midway, below right]    {$\cal{J}^-$}(0bot) --node[midway, above,sloped] {$r=0$}(0top) -- cycle;
\end{tikzpicture}
\caption{ \label{fig:CDs} Conformal diagrams of the solution found in section \ref{sec:Solution} in the three cases $M>M_{\min}$, $M=M_{\min}$, $M<M_{\min}$.}
\end{figure*}
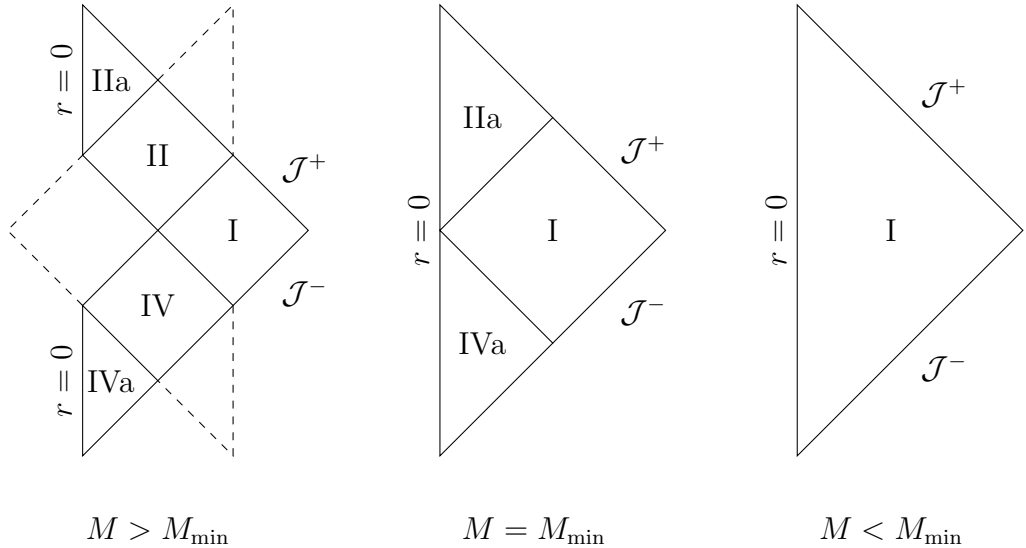
\begin{description}
\item[$M>M_{\min} :$] The eternal black hole solution is covered by the regions $\textup{I}$, $\textup{II}$, $\textup{IIa}$. Regions $\textup{I}$, $\textup{IV}$, $\textup{IVa}$ describe the time reversed white hole solution. Dashed lines indicate the mirror symmetric extension.

\item[$M=M_{\min} :$] The outer and inner horizon coincide and the regions $\textup{II}$ and $\textup{IV}$ shrink to a single horizon $r_+=r_-=r_*$.
\item[$M<M_{\min} :$] No horizon occurs and the causal structure is just like Minkowski spacetime. Close to $r=0$ the solution approaches a static de Sitter metric replacing the singularity.
\end{description}

\clearpage
\pagestyle{empty}
\clearpage
\pagestyle{plain}

\subsection{Electric charge}
The Reissner-Nordstr\"om metric in Schwarzschild coordinates takes the form (\ref{Schwspfl}), where now
\begin{equation}
    1-a^2 = 1-\frac{2M}{r}+\frac{Q^2}{r^2}.
\end{equation}
Using the same transformation to Lema\^itre coordinates as in \cite{BlackHoleRemnants}, we find that in this case they can only cover the part $r> \frac{Q^2}{2M}$ of this spacetime which contains both horizons but not the singularity. The expressions for $a(x)$ and $b(x)$ in (\ref{mansatz}) can be found from the relation
\begin{equation}
    r(x) = \frac{Q^2}{2M}\left(\theta(\bar{x})+ \theta^{-1}(\bar{x}) -1\right),
\end{equation}
where
\begin{equation}
    \theta^3(\bar{x}) = 1+2\bar{x}^2\left(1+ \sqrt{1+\bar{x}^{-2}}\right),\qquad \bar{x}:= \frac{3M^2\, x}{Q^3}.
\end{equation}
More generally, also the synchronous coordinates associated to the solution $\xi^{(c>1)}_+$ (cf. section \ref{OnGen}) cover only the region
\begin{equation}
    r>\frac{M}{c-1}\left(\sqrt{1+(c-1)\frac{Q^2}{M^2}}-1\right).
\end{equation}
Hence, even though one can obtain synchronous coordinates covering regions arbitrarily close to $r=0$, there is no global synchronous coordinate system covering the Reissner-Nordstr\"om metric. The constraint (\ref{constraint}) does not have a global solution and no synchronous Cauchy hypersurface exists. This can be taken as an indicator of the pathologies associated to the unstable interior of this spacetime which exhibits a Cauchy horizon \cite{FrolovBH}.

Searching for a modification of this metric in our modified theory of gravity, we, however, have to care only about matching this GR solution in the low curvature limit in the exterior. 
Since this metric still respects spherical symmetry and the Killing vector field $\partial/\partial t$, the ansatz (\ref{mansatz}) is general enough to cover also modifications of charged black holes. 
Note that for the Reissner-Nordstr\"om metric in Lema\^itre coordinates the trace of extrinsic curvature expanded around $x=0$ reads
\begin{equation}
    \kappa = -\frac{1}{x} -\frac{16M^4}{3Q^2} x + \mathcal{O}(x^3).
\end{equation}
It is hence singular at the point $x=0$ corresponding to $r=Q^2/2M$ already before the actual curvature singularity\footnote{$R^{\mu\nu}R_{\mu\nu} = 4Q^4/r^8$, $R^{\mu\nu\alpha\beta}R_{\mu\nu\alpha\beta} = 8\left(Q^4+6\left(Q^2-Mr\right)^2\right)/r^8$} at $r=0$ is reached. The modification must hence anyway take over well before this point is reached.

Since the metric is of the form (\ref{Schwspfl}), we know already that the Reissner-Nordstr\"om solution is spatially flat in Lema{\^i}tre coordinates, just like the Schwarzschild solution.
The main difference in going from the uncharged to the charged case is that now we are no longer looking for vacuum solutions but for \textit{electro}vacuum solutions.
In the exterior we expect the static observers defined by the Killing vector field $\partial/\partial t$ to observe only an electric but no magnetic field. Moreover, due to spherical symmetry, this electric field should only depend on the Schwarzschild $r$-coordinate. We hence use the ansatz
\begin{equation}
    A_\mu \mathrm{d}x^\mu = \Phi(r(x)) \mathrm{d}t = \Phi(x) \frac{\mathrm{d}T-a^2\mathrm{d}R}{1-a^2}
\end{equation}
for the electromagnetic potential 1-form. In Lema{\^i}tre coordinates, its components with raised index are
\begin{equation}
    A^T =  A^R = \frac{\Phi(x) }{1-a^2}. 
\end{equation}
It follows that the only non-vanishing components of the Faraday tensor are
\begin{equation}
    F^{TR}=-F^{RT}=\frac{2}{a^2}\frac{\partial}{\partial x} A^T.
\end{equation}
The vacuum Maxwell equation amounts to
\begin{equation}
    F^{T\mu}_{;\mu} = \frac{1}{ab^2} \frac{\partial}{\partial R} \left(ab^2 F^{TR}\right) = 0,
\end{equation}
from which it follows that
\begin{equation}
    F^{TR}= \frac{Q}{ab^2} \qquad \textup{and} \qquad \frac{\partial}{\partial x} A^T = 2Q \frac{a}{b^2},
\end{equation}
where a constant of integration corresponding to charge was fixed.
The energy momentum tensor of the electromagnetic field is
\begin{equation}
    T^\alpha_\beta = \frac{1}{4\pi} \left(F^{\alpha\mu}F_{\mu\beta}-\frac{1}{4}F^{\mu\nu}F_{\mu\nu}\, \delta^\alpha_\beta\right)  = \frac{Q^2}{8\pi b^4}\,  \mathrm{diag}\left(-1,-1,1,1 \right ).
\end{equation}

Inserting this into the modified Einstein equation, the temporal equation takes the form
\begin{equation}
    \frac{1}{3}\left(  f-2\kappa f^{\prime}\right)  \kappa^{2}-\Lambda
+\kappa\Lambda^{\prime}-\frac{1}{2}\left(  f+\kappa f^{\prime}\right)
\tilde{\kappa}_{b}^{a}\tilde{\kappa}_{a}^{b}=\gamma^{RR} \left(\kappa^\vartheta_{\vartheta}\right)^2 - \gamma^{\vartheta\vartheta} -\frac{Q^2}{b^4},
\label{chbh00}
\end{equation}
and the analogue of (\ref{speqsub}) becomes
\begin{equation}
   \frac{1}{ab^2} \left(ab^2 f \left(\frac{b'}{b}-\frac{a'}{a}\right)\right)' = \frac{b}{2b'}\left(\gamma^{RR}(\kappa^\vartheta_\vartheta)^2 - \gamma^{\vartheta\vartheta}\right)' - \frac{2 Q^2}{b^4}.
   \label{chbhspfl00}
\end{equation}
Assuming spatial flatness, i.e. $a = b'$, this equation has the first integral
\begin{equation}
    \frac{b'}{b}-\frac{a'}{a} = \frac{1}{ab^2 f} \left(3M +\frac{2 Q^2}{b}\right).
\end{equation}
Moreover, in this case
\begin{equation}
    A^T = - \frac{2Q}{b},
\end{equation}
and the temporal equation (\ref{chbh00}) becomes
\begin{equation}
\kappa^2\left(f-2\kappa f'\right)-3\left(\Lambda - \kappa \Lambda'\right)= \left(f+\kappa f'\right) \left(\frac{3M+ 2 Q^2/b}{fu}\right)^2 - \frac{3Q^2}{b^4}.
\label{BH00sfc}
\end{equation}
We see that the new terms coming from charge can become dominant only close to $b=r\to0$. They remain negligible until well after the modifications have taken over at $r_*$, provided that
\begin{equation}
    \left(M^4 M_{\min}^2\right)^{1/3}\gg Q^2.
\end{equation}
Since we can expect the Schwinger effect to discharge a black hole much faster than the timescales of Hawking radiation, this condition should be satisfied by realistic black holes close to the end of their evolution. This suggests that our conclusions drawn for the fate of evaporating uncharged black holes in \cite{BlackHoleRemnants} should remain valid also in the charged case.

Noting the sign of the charge contribution to (\ref{chbhspfl00}), there is the possibility that charge can lead to a bounce even without a spatial curvature dependent potential. As we can check explicitly by inserting the modified solution from section \ref{sec:Solution}, the right hand side of (\ref{chbhspfl00}) in this case exhibits a zero. Such a bounce would also prevent a blow up of the electromagnetic field energy at $r=0$. Of course, a rigorous verification of this speculation would require a more extensive analysis beyond the scope of this paper.

\subsection{First order rotation}
To first order in angular momentum $J$ the Kerr metric in Boyer-Lindquist coordinates reads \cite{FrolovBH}
\begin{equation}
    \mathrm{d}s^2 = \left(1-\frac{r_g}{r}\right)\mathrm{d}t^2-\frac{\mathrm{d}t^2}{\left(1-\frac{r_g}{r}\right)}-r^2\mathrm{d}\Omega^2+\frac{4J}{r}\sin^2\vartheta\,\mathrm{d}t\mathrm{d}\varphi
\end{equation}
and is still spherically symmetric.
In the Lema\^itre coordinates
\begin{equation*}
\mathrm{d}T = \mathrm{d}t + \sqrt{\frac{r_g}{r}}\left ( 1-\frac{r_g}{r} \right )^{-1}\mathrm{d}r, \qquad
\mathrm{d}R = \mathrm{d}t + \sqrt{\frac{r}{r_g}}\left ( 1-\frac{r_g}{r} \right )^{-1}\mathrm{d}r,
\end{equation*}
it becomes
\begin{equation}
    \mathrm{d}s^2 = \mathrm{d}T^2-a^2(x)\mathrm{d}R^2-b^2(x)\mathrm{d}\Omega^2+  \frac{2 j\,\omega(x) \sin^2\vartheta}{1-a^2(x)} \left(\mathrm{d}T-a^2(x)\mathrm{d}R\right)\mathrm{d}\varphi, \label{eq-ansatz}
\end{equation}
where $x=R-T$, $j:= J/M^2$ and
\begin{equation}
    a(x) = \left(\frac{x}{x_{+}}\right)^{-1/3}, \quad b(x) = \left(\frac{x}{x_{+}}\right)^{2/3} r_g, \quad \omega(x) =  M\left(\frac{x}{x_{+}}\right)^{-2/3}.
    \label{SchwLemomeg}
\end{equation}
Let us take (\ref{eq-ansatz}) as an ansatz where we consider $a$, $b$ and $\omega$ as independent functions.
Note that even though these coordinates are not synchronous, to first order in the angular momentum it still holds that
\begin{equation}
    g^{\mu\nu}T_{,\mu}T_{,\nu}= 1+\mathcal{O}\left(j^2\right).
\end{equation}
Making the expansion
\begin{equation}
    \phi = T + j\, \phi_0(x) + \mathcal{O}\left(j^2\right),
\end{equation}
where $\phi_0(x)$ should depend on $x$ to preserve spherical symmetry, we find the condition $\phi_0' =0$. Hence, in first order perturbation theory we can still use the approximate solution
\begin{equation}
    \phi = T + \mathcal{O}\left(j^2\right).
\end{equation}
However, now  $\phi^{,\varphi} = g^{T\varphi}\neq0$ and
\begin{equation}
    \phi_{;T\varphi}= \frac{j \,\omega \sin^2\vartheta}{(1-a^2)}\frac{b'}{b} + \mathcal{O}\left(j^3\right)\neq 0.
\end{equation}
Hence, we cannot directly use the components of the modified Einstein equation that were derived in the synchronous coordinates above. Instead, we have to expand the full equation (\ref{mEE}) in $j$.
Expanding the Ricci tensor
\begin{equation}
    R_{\mu\nu} = {^{(0)}\!R_{\mu\nu}} + j\, {^{(1)}\!R_{\mu\nu}} + \mathcal{O}\left(j^2\right),
\end{equation}
we find that the only new non-vanishing contibutions to first order in angular momentum are given by
\begin{equation}
    {^{(1)}\!R^T_{\varphi}} =  {^{(1)}\!R^R_{\varphi}}=\frac{\sin^2\vartheta}{a^2}\left[\frac{1}{2}a\left(\frac{\omega'}{a}\right)'-\omega\left(\frac{b''}{b} +\left(\frac{b'}{b}\right)^2-\frac{a'}{a}\frac{b'}{b}\right)\right].
\end{equation}
In first order perturbation theory it still holds that
\begin{equation}
    \phi^{;T}_{\:T} = 0, \quad \phi^{;R}_{\:R} = -\frac{a'}{a}, \quad \phi^{;\vartheta}_{\:\vartheta} = \phi^{;\varphi}_{\:\varphi} = -\frac{b'}{b}, \quad \Box \phi = - \frac{a'}{a}-2\frac{b'}{b},
\end{equation}
where corrections would appear in order $\mathcal{O}(j^2)$. Moreover,
\begin{equation}
     \phi^{;T}_{\:\varphi} = 0, \quad \phi^{;T\;\:T}_{\:\:\:\varphi} = 0, \quad
    \phi^{;R}_{\:\varphi}  = \frac{j \sin^2\vartheta}{2a^2}\, b^2\left(\frac{\omega}{b^2}\right)'
    \label{foRcovdercomp}
\end{equation}
where corrections would appear in order $\mathcal{O}\left(j^3\right)$.

The functions $a$ and $b$ are hence still determined by the zeroth order equations (\ref{BH00}), (\ref{speqsub}). 
The function $\omega$ has to be obtained from one of the new off-diagonal modified Einstein equations, e.g. the $T-\varphi$ equation which, making use of (\ref{foRcovdercomp}), becomes
\begin{equation}
     f R^T_\varphi +\phi^{;R}_{\:\varphi} f_{,R} =0.
\end{equation}
 More explicitely, we have to determine $\omega$ from the equation
\begin{equation}
   f\left( a\left(\frac{\omega'}{a}\right)'-2\left(\frac{b''}{b} +\left(\frac{b'}{b}\right)^2-\frac{a'}{a}\frac{b'}{b}\right)\omega \right) + f_{,R}\,b^2\left(\frac{\omega}{b^2}\right)'= 0.
\end{equation}
Assuming spatial flatness of the slices $T=const.$ (i.e. $a=\pm b'$) this simplifies to
\begin{equation}
   f\left( a\left(\frac{\omega'}{a}\right)'-2\left(\frac{b'}{b}\right)^2\omega \right) + f_{,R}\,b^2\left(\frac{\omega}{b^2}\right)'= 0.
   \label{omegaODE}
\end{equation}
It is easy to see that one solution of this linear ODE is given by $\omega\propto b^2$. However, this solution does not agree with the asymptotic solution (\ref{SchwLemomeg}). Multiplying (\ref{omegaODE}) by $b^2/a$ and using (\ref{speqsub}) in the spatially flat case, it is straightforward to verify that another independent solution is given by
\begin{equation}
    \omega = M a^2,
\end{equation}
where the constant of integration was already fixed to match (\ref{SchwLemomeg}). Note that this identity for the GR solution continues to hold in the modification in the spatially flat case.

It follows that the frame dragging function $\omega$ assumes a maximum at the same location as $a$  at $r=r_*$ and after that decreases until it vanishes at $r=0$ according to $\omega \propto r^2$. Since $\omega$ is bounded by a maximum value, we can expect that for a small enough value of $j\ll 1$ our perturbative analysis is justified for the whole range of $x$. Moreover, this suggests that the spacetime structure of a rotating black hole close to $r=0$ is in fact not different from the non-rotating case.

Note that in first order perturbation theory the norm of the Killing vector field (and hence the location of horizons) is still given by $a^2-1$. Moreover, for the surface gravity it holds that
\begin{equation}
    g_s = -a'(x_+) + \mathcal{O}(j^2).
\end{equation}
This shows that our above conclusions are robust even for slowly rotating black holes.

\clearpage
\section{Conclusions}
The introduction of the mimetic field $\phi$ allowed us to find a remarkably simple high-curvature modification of GR, where a scale dependence of gravitational and cosmological constant can be implemented covariantly. We found that $\Box \phi$ is the unique measure of curvature on which the gravitational constant can depend such that the resulting modified Einstein equation is still second order in time. This modified theory of gravity hence does not exhibit any additional degrees of freedom except that the conformal degree of freedom of the metric becomes dynamical. 

As a first application, we found that the most natural class of modified Friedmann universes arising from this theory generically feature a de Sitter-like initial state replacing the Big Bang singularity. To resolve also the anisotropic Kasner singularity in the same way, we found that we have to require ``asymptotic freedom'' of gravity, i.e. the vanishing of the gravitational constant at limiting curvature. 

Taking on the task of singularity resolution in general, spatially non-flat spacetimes, it is clear that this is too much to ask of a theory where only the conformal degree of freedom is modified. Gratifyingly, we found that the mimetic field also permits to introduce in a covariant manner a potential depending on spatial curvature. In fact, adding such higher order terms to the action could even improve the renormalizability of gravity, along the lines of Ho\v{r}ava gravity.
We showed that in spatially non-flat Friedmann and certain Bianchi universes a simple power law potential is already enough to replace the singularity with a bounce.

In application to non-rotating black holes, we found that our modification of GR generically leads to a lower bound on the black hole mass. Minimal black holes have vanishing Hawking temperature and the final product of black hole evaporation is hence a stable remnant of minimal mass.
Moreover, we found that this result is also robust when putting small amounts of charge or angular momentum.
An inner horizon is already present in the non-rotating, uncharged case and the causal structure resembles those of Reissner-Nordstr\"om and Kerr, except that the region inside the inner horizon is replaced with a static de Sitter patch. 
Furthermore, since the mere assumption of existence of a global solution to the mimetic constraint already implies stable causality, we expect no Cauchy horizon in the interior even for arbitrary charge and rotation. Hence the instabilities present in Reissner-Nordstr\"om and Kerr solutions could be cured in such a modification. 

\clearpage
\section*{Appendix}
\addcontentsline{toc}{section}{Appendix}

\subsection*{A: Synchronous coordinates.}
\addcontentsline{toc}{subsection}{A: Synchronous coordinates}
Variation of the action (\ref{action}) with respect to the Lagrange multiplier $\lambda$ yields the constraint (\ref{constraint})
\begin{equation*}
    g^{\mu\nu}\phi_{,\mu}\phi_{,\nu} = 1.
\end{equation*}
We will see that the existence of a global scalar field with this porperty already has some far reaching consequences, e.g. on the causal structure of admissible spacetimes.\footnote{For example, as shown in \cite{HawkingEllis}, the existence of a function whose gradient is everywhere time-like implies stable causality.} Taking a covariant derivative of this equation shows that
\begin{equation}
    \phi^{,\mu}\nabla_{\mu}\phi^{,\nu} = \frac{1}{2}\, \nabla^{\nu}\left(\phi^{,\mu}\phi_{,\mu}\right)=0,
    \label{geodesicness}
\end{equation}
and hence the vector field $\phi^{,\mu}$ is tangent to a congruence of timelike geodesics. Through every point of a hypersurface of constant $\phi$ passes a unique geodesic in the congruence. Choosing coordinates\footnote{More generally: an atlas} $x^a$ on some initial $3-$hypersurface $\{\phi=\phi_i=const.\}$ then defines coordinates on any other hypersurface $\{\phi = \phi_0= const.\}$ by traveling along these unique geodesics. Since the congruence is hypersurface orthogonal and its normal vector field $\phi^{,\mu}$ has unit norm,  $(t:=\phi,x^a)$ defines a synchronous coordinate system in which the metric takes the form \cite{Landau}
\begin{equation*}
    \textup{d}s^{2}=\textup{d}t^{2}-\gamma_{ab}\textup{d}x^{a}\textup{d}x^{b},
\end{equation*}
where latin indices run over spatial coordinates. The whole spacetime is sliced into into spatial hypersurfaces $\{\phi = \textup{const.}\}$ with extrinsic curvature
\begin{equation*}
    \kappa_{ab} = \frac{1}{2} \frac{\partial }{\partial t} \gamma_{ab}, \qquad \kappa^{ab} := \gamma^{ac}\gamma^{bd}\kappa_{cd} = -\frac{1}{2} \frac{\partial }{\partial t} \gamma^{ab}, \qquad \kappa := \gamma^{ab}\kappa_{ab} =  \frac{\partial }{\partial t} \ln \sqrt{\gamma}
\end{equation*}
and metric determinant $\gamma = \det{\gamma_{ab}} = - \det{g_{\mu\nu}}$.
In this splitting the non-vanishing connection coefficients are given by
\begin{gather*}
    \Gamma^{0}_{ab}= \kappa_{ab},\qquad\Gamma^{a}_{0b}=\kappa^a_{b} := \gamma^{ac}\kappa_{cb}, \qquad \Gamma^{a}_{bc} =\lambda ^{a}_{bc},
\end{gather*}
where $\lambda ^{a}_{bc}$ are the connection coefficients of the Levi-Civita connection $D$ belonging to the Riemannian $3-$metric $\gamma_{ab}$. 
Note that 
\begin{equation}
    \phi_{;0\alpha} = 0, \qquad \phi_{;ab} = - \kappa_{ab},
\end{equation}
and the expansion and shear of the geodesic congruence $\phi^{,\mu}$ are given by $\Box\phi = \kappa$ and $\tilde{\kappa}^a_b := \kappa^a_b-\tfrac{1}{3}\kappa \delta^a_b$, respectively.

The covariant 4-divergence of a vector $X^{\mu}$ is given by
\begin{equation*}
     \nabla_\mu X^{\mu}  = \partial_0 X^0 + \kappa X^0 +D_a X^{a} ={ \tfrac{1}{\sqrt{\gamma}} }\partial_0\left (\sqrt{\gamma } X^0   \right )+D_a X^{a}
\end{equation*}
and the d'Alembertian of a scalar $S$ is
\begin{equation*}
     \Box S ={ \tfrac{1}{\sqrt{\gamma}} }\,\partial_0\!\left (\sqrt{\gamma }\, {\dot{S} } \right ) - \Delta S = \ddot{S} +\kappa\dot{S}- \Delta S
\end{equation*}
where the dot denotes $t$-derivatives and $\Delta$ is the Laplacian belonging to the Riemannian metric $\gamma$.

The non-vanishing components of the four-dimensional Riemann tensor in this splitting are determined by
\begin{gather*}
R_{\: abc}^{0}    =\kappa_{ac|b}-\kappa_{ab|c} \qquad \qquad
R_{\: a0b}^{0}    =\dot{\kappa}_{ab}-\kappa_{a}^{c}\kappa_{bc}\\
R_{\: abc}^{d}    = {^{3}\!R_{\: abc}^{d}}+\kappa_{b}^{d}\kappa_{ca}
-\kappa_{c}^{d}\kappa_{ab}\phantom{\frac{1}{1}}
\end{gather*}
where ${^{3}\!R_{\: abc}^{d}}$ is the Riemann tensor of the spatial metric $\gamma_{ab}$ and the notation $\%_{|b}:= D_b\%$ was used.
We find the useful identity
\begin{equation}
    \phi^{,\mu}\phi_{,\nu}R^{\nu}_{\alpha\mu\beta} = -\phi^{,\mu}\nabla_{\mu}\left(\phi_{;\alpha\beta}\right) - \phi^{;\mu}_{\;\alpha}\phi_{;\mu\beta}.
    \label{Riemannid}
\end{equation}
The Ricci tensor $R^{\mu}_{\nu}$ splits into extrinsic and intrinsic curvature as
\begin{gather*}
    R^{0}_{0} = R_{00} =  -\dot{\kappa}-\kappa^{ab}\kappa_{ab}\qquad\qquad
R^{0}_{a} = R_{0a} = \kappa^{b}_{a|b} - \kappa_{,a}\\
    -R^{a}_{b} = \gamma^{ac} R_{cb} = \tfrac{1}{\sqrt{\gamma }}\partial_0 \left (\sqrt{\gamma } \kappa^{a}_{b} \right ) +  {^3\!}R^{a}_{b}. \phantom{\frac{1}{1}}
\end{gather*}
The Ricci scalar is thus given by
\begin{equation*}
    {-\!R} = 2\dot{\kappa} +\kappa^2+\kappa^{ab}\kappa_{ab} + {^3\!}R.
\end{equation*}
The $0-0$ component of the Einstein tensor $G^{\mu}_{\nu} = R^{\mu}_{\nu}-\frac{1}{2} R \delta^{\mu}_{\nu} $ is hence
\begin{equation*}
   G_{00}= G^{00}= G^{0}_{0}=  \tfrac{1}{2}\left ( \kappa^2-\kappa^{ab}\kappa_{ab} + {^3\!}R \right ).
\end{equation*}
This allows to isolate the spatial curvature scalar as
\begin{equation*}
    {^3\!}R = 2G_{\mu\nu}\phi^{,\mu}\phi^{,\nu}-(\Box\phi)^2+\phi^{;\mu\nu}\phi_{;\mu\nu}.
\end{equation*}
For evaluation of the modified Einstein equation in the synchronous frame it will be useful to note that
\begin{equation}
    \nabla_\mu\left(\phi^{;\alpha}_{\;\beta}\tilde{f} \phi^{,\mu}\right) =  \tfrac{1}{\sqrt{\gamma }}\partial_0 \left (\sqrt{\gamma } \tilde{f} \kappa^{a}_{b} \right )  \delta^\alpha_a \delta_\beta^b.
\end{equation}

\subsubsection*{A causal completeness condition}
In this section we will find a sufficient (but not necessary) condition for causal geodesic completeness of a metric of the form (\ref{synch}). We follow mainly the steps taken in \cite{ChoquetBruhat}, applied to the $3+1$ splitting at hand.
Consider the velocity vector $u^\mu$ of a geodesic parametrized by an affine parameter $s$,
\begin{equation*}
    u^0 = \frac{\textup{dt}}{\textup{ds}},\qquad u^a = \frac{\textup{d}x^a}{\textup{ds}} = u^0\frac{\textup{d}x^a}{\textup{dt}} =: u^0v^a.
\end{equation*}
The temporal component of the geodesic equation reads
\begin{equation*}
   0 = u^\mu \nabla_\mu u^0 =  \frac{\textup{d}}{\textup{ds}} u^0 + \kappa_{ab}u^a u^b = u_0^2 \left(\frac{1}{u_0}\frac{\textup{d}}{\textup{dt}} u^0 + \kappa_{ab}v^a v^b \right)
\end{equation*}
and we can integrate to find
\begin{equation}
    \ln u^0 = -\int\!\mathrm{d}t\, \kappa_{ab}v^a v^b. 
    \label{eq-geodcomp}
\end{equation}
Timelike geodesics with $u^\mu =\phi^{,\mu}$ describe ``comoving'' observers with $v^a=0$. They are freely falling and the synchronous time $t$ measures proper time for these observers.
Their affine parameter extension is hence infinite, if $t$ can be extended to the range $(-\infty,\infty) $. 
Assuming this to be the case, the affine parameter extension of a general causal geodesic is given by
\begin{equation}
    \int \textup{d} s = \int_{-\infty}^{\infty} \frac{\textup{d} t}{u^0},
    \label{eq-paramintegral}
\end{equation}
and thus it is future resp. past complete, if this integral diverges at $t\to \infty$ resp. $t\to -\infty$. Using the Cauchy-Schwarz inequality for the scalar product $\left \langle A ,B\right \rangle := \gamma^{ab}\gamma^{cd} A_{ac} B_{bd}$ of spatial tensors $A$ and $B$ on the right hand side of (\ref{eq-geodcomp}), we see that
\begin{equation*}
    \ln u^0 \leq \int\!\mathrm{d}t\, \sqrt{\kappa^{ab}\kappa_{ab}}(v^c v_c) \leq \int\!\mathrm{d}t\, \sqrt{\kappa^{ab}\kappa_{ab}},
\end{equation*}
where the second inequality is an equality for light-like geodesics. It hence follows that if 
\begin{equation}
    \int_{-\infty}^{\infty}\!\mathrm{d}t\, \sqrt{\kappa^{ab}\kappa_{ab}} < \infty,
    \label{eq-kabkabbound}
\end{equation}
then $1/u^0$ will be uniformly bounded from $0$ and hence all causal geodesics are both past- and future complete. 
Note that also the weaker condition
\begin{equation}
    \sqrt{\kappa^{ab}\kappa_{ab}} \leq\frac{1}{\left|t\right|} \quad \textup{asymptotically as} \quad t\to\pm\infty,
    \label{eq-kabkabbound2}
\end{equation}
suffices to have $u^0 \leq\left|t\right|$ and hence logarithmic divergence of (\ref{eq-paramintegral}). Thus also (\ref{eq-kabkabbound2}) is a sufficient condition for causal completeness.

\subsection*{B: Explicit calculations in the variation of the action}
\addcontentsline{toc}{subsection}{B: Explicit calculations in the variation of the action}
\paragraph{Variation with respect to the mimetic field.} Let us vary (\ref{action}) with respect to $\phi$. To this end, calculate
\begin{align}
    \frac{ \delta\mathcal{L} }{\delta \Box\phi} &= f'\left (R+2G_{\mu\nu}\phi^{,\mu}\phi^{,\nu} -\left ( \Box \phi \right )^2 + \phi^{;\mu\nu}\phi_{;\mu\nu}   \right )-2(f-1+h')\Box\phi+2\Lambda' \phantom{\frac{1}{1}}\nonumber\\ 
    &\doteq  -2\left [(\Tilde{f}-h')_{,\alpha}\phi^{,\alpha}+\Tilde{f}\,\Box\phi+\tfrac{1}{2}f'\left ( (\Box\phi)^2+ \phi^{;\mu\nu}\phi_{;\mu\nu}   \right )-\Lambda' \right ] \nonumber \\ &=: -2\left [( \Tilde{f}\phi^{,\alpha} )_{;\alpha}+\Tilde{Z} \right ] \label{dLdBoxphi}
\end{align}
where we introduced the useful notations
\begin{equation*}
    \Tilde{f} := f-1+h',\qquad Z:=\tfrac{1}{2}f'\left((\Box\phi)^2+\phi^{;\mu\nu}\phi_{;\mu\nu} \right) - \Lambda', \qquad \Tilde{Z} := Z - \phi^{,\alpha}h'_{,\alpha}
\end{equation*}
and $\doteq$ means equality if the constraint (\ref{constraint}) is satisfied. It follows that
\begin{align*}
\tfrac{1}{2}\,\delta_\phi\mathcal{L} &=-\left [ ( \Tilde{f}\phi^{,\mu} )_{;\mu}+\Tilde{Z} \right ] \Box \delta\phi+ \Tilde{f}\left ( 2G_{\mu\nu}\phi^{,\mu}\,\delta \phi^{,\nu}+ \phi^{;\mu\nu}\delta\phi_{;\mu\nu} \right ) \phantom{\frac{1}{1}}\\ 
 &= -\left [( \Tilde{f}\phi^{,\mu} )_{;\mu}+\Tilde{Z} \right ]
\delta\phi^{;\nu}_{\:\nu}+ \Tilde{f}\left (\phi^{;\mu}_{\:\nu}\,\delta\phi^{,\nu}   \right )_{;\mu}   -\Tilde{f}\left [\phi^{;\mu}_{\:\nu\mu}- 2G_{\mu\nu}\phi^{,\mu}  \right ]\,\delta\phi^{,\nu}  \phantom{\frac{1}{1}}
\end{align*}
Thus the variation of (\ref{action}) yields
\begin{align*}
-8\pi\delta_\phi S &=\int\! \textup{d}^4x \sqrt{-g}\,\delta\phi^{,\nu}\left \{( \Tilde{f}\phi^{,\mu}  )_{;\mu\nu} - ( \Tilde{f} \phi^{;\mu}_{\:\nu} )_{;\mu} +\Tilde{Z}_{;\nu}+\Tilde{f}2G_{\mu\nu}\phi^{,\mu}-\lambda \phi_{,\nu}\right \}  \\
&=\int\! \textup{d}^4x \sqrt{-g}\,\delta\phi^{,\nu}\left \{ ( \Tilde{f}_{,\nu}\,\phi^{,\mu} )_{;\mu}+  \Tilde{Z}_{,\nu}+\Tilde{f}\left (2G_{\mu\nu}-R_{\mu\nu} \right )\phi^{,\mu} -\lambda \phi_{,\nu}\right \}  
\end{align*}
where covariant partial integration and the commutator of covariant derivatives were used. Here and in the following section we ignore boundary terms in the variation. Integrating by parts once again, we find the equation of motion
\begin{equation*}
    \nabla_\nu\left [( \lambda +\Tilde{f}R  )\phi^{,\nu} - ( \Tilde{f}^{,\nu}\,\phi^{,\mu} )_{;\mu} -\Tilde{Z}^{,\nu}-\Tilde{f}R^{\mu\nu}\phi_{,\mu}\right ]= 0
\end{equation*}
which will be used to determine $\lambda$.

\clearpage
\paragraph{Variation with respect to the metric.}
Next we have to vary (\ref{action}) with respect to $g_{\mu\nu}$. 
In the course of this undertaking the following identities for the varations of the metric determinant, connection coefficients and Ricci tensor will be put to good use:
\begin{equation*}
    \delta \sqrt{-g} = -\frac{1}{2}\sqrt{-g}\, g_{\mu\nu} \delta g^{\mu\nu},\qquad 
\delta {\Gamma} ^{\lambda }_{\mu\nu } = - g_{\alpha(\mu}\nabla_{\nu)}\delta g^{\alpha\lambda} + \tfrac{1}{2} g_{\mu\alpha}g_{\nu\beta}\nabla^{\lambda}\delta g^{\alpha\beta}
\end{equation*}
\begin{equation*}
    \delta R_{\mu\nu} = \nabla_{\lambda}\delta \Gamma^{\lambda}_{\mu\nu}-\nabla_{\nu}\delta \Gamma^{\lambda}_{\lambda\mu}
\end{equation*}
Combining the latter two yields
\begin{equation*}
    \delta R _{\mu\nu } = \tfrac{1}{2}\left [ g_{\mu\alpha}g_{\nu\beta}\Box+g_{\alpha\beta}\nabla_{\nu}\nabla_{\mu} -g_{\mu\beta}\nabla_{\alpha}\nabla_{\nu}  -g_{\nu\beta}\nabla_{\alpha}\nabla_{\mu}\right ]\delta g^{\alpha\beta}.
\end{equation*}
In the variation of the usual Einstein action one only encounters the term
\begin{equation*}
    g^{\mu\nu}\delta R _{\mu\nu } = \left (g_{\mu\nu} \,\Box  - \nabla_\mu \nabla_\nu  \right ) \delta g^{\mu \nu },
\end{equation*}
which turns out to be a total covariant derivative, provided that it appears with a constant prefactor. 
Using, in a first step,
\begin{equation*}
    \delta_g\left ( 2 G_{\mu\nu}\phi^{,\mu}\phi^{,\nu} \right ) =2 \delta R_{\mu\nu}\phi^{,\mu}\phi^{,\nu} +4R_{\alpha\mu}\phi^{,\alpha}\phi_{,\nu}\delta g^{\mu\nu} -\delta R-R\phi_{,\mu}\phi_{,\nu}\delta g^{\mu\nu},
\end{equation*}
the expression $\delta \mathcal{L}/\delta \Box \phi$ from (\ref{dLdBoxphi}) and ignoring boundary terms we find that 
\begin{align*}
   -16\pi\, &  \delta_g S=  \phantom{\frac{1}{1}} \\
  \int\!\textup{d}^4x&\,\sqrt{-g} \Bigg\{  \left [ R_{\mu\nu}-\frac{1}{2} \mathcal{L}\, g_{\mu\nu} +4\Tilde{f}\phi^{,\alpha}R_{\alpha(\mu}\phi_{,\nu)}-\left(\lambda+\Tilde{f}R \right)\phi_{,\mu}\,\phi_{,\nu}  \right ] \,\delta g^{\mu\nu}  \\
    & -\left[(\Tilde{f}\phi^{,\alpha})_{;\alpha} + \Tilde{Z}\right ]\,\underbrace{2\,\delta_g \Box\phi }_{\textsc{\romannumeral 1}}+\Tilde{f}\underbrace{2\,\phi^{,\mu}\phi^{,\nu}\delta R_{\mu\nu}}_{\textsc{\romannumeral 2}}
    + \Tilde{f}\underbrace{\delta_g \left(\phi^{;\mu\nu}\phi_{;\mu\nu}\right)}_{\textsc{\romannumeral 3}} 
    - h' \underbrace{\,\delta R\phantom{_\mu} }_{\textsc{\romannumeral 4}}  \Bigg \}
\end{align*}
The modified Einstein equation hence reads
\begin{align*}
G_{\mu\nu} - \Lambda g_{\mu\nu}&-\tfrac{1}{2}g_{\mu\nu}\left [(f-1)(R +{^3\!R}) +h\right ]+ 4\phi^{,\alpha}\phi_{(,\mu}R_{\nu)\alpha} +\dots \phantom{\frac{1}{1}}\\
& \dots-T^{\textsc{\romannumeral1}}_{\mu\nu}+T^{\textsc{\romannumeral2}}_{\mu\nu}+T^{\textsc{\romannumeral3}}_{\mu\nu}-T^{\textsc{\romannumeral4}}_{\mu\nu}= (\lambda + \Tilde{f}R) \phi_{,\mu}\phi_{,\nu}  + 8 \pi T_{\mu\nu}^{\textup{(m)}},\phantom{\frac{1}{1}}
\end{align*}
where we still have to figure out the contribution of the terms $\textsc{\romannumeral1}-\textsc{\romannumeral4}$.

\newpage
Starting with term $\textsc{\romannumeral1}$, we first have to calculate
\begin{equation*}
2\,\delta_g \Box\phi = -\phi^{,\mu}\nabla_\mu\left ( g_{\alpha \beta }\delta g^{\alpha \beta }  \right ) + 2\nabla_\mu\left (\delta g^{\mu\nu}\phi_{,\nu}  \right )
\end{equation*}
where only the variation of the metric determinant and the identity 
\begin{equation*}
    {\Gamma} ^{\nu }_{\nu\mu} = \frac{1}{\sqrt{-g}}\partial_{\mu}\sqrt{-g}
\end{equation*}
were used. The contribution to the variation of the action of a term like $\textsc{\romannumeral1}$ multiplied by an arbitrary spacetime function $\mathcal{F}$ is thus
\begin{align*}
\int \textup{d}^4x\sqrt{-g} \,\mathcal{F}\,2\,\delta_g \Box \phi  &=\int \textup{d}^4x\sqrt{-g}\,\mathcal{F} \left (-\nabla_\alpha\left ( g_{\mu\nu }\delta g^{\mu\nu }  \right )\phi^{,\alpha} + 2\nabla_\mu\left (\delta g^{\mu\nu}\phi_{,\nu}  \right )   \right ) \\
&= \int \textup{d}^4x\sqrt{-g}\left (g_{\mu\nu } (\mathcal{F}\phi^{,\alpha})_{;\alpha} -2\mathcal{F}_{(,\mu}  \phi_{,\nu)}  \right )\delta g^{\mu\nu}
\end{align*} 
where covariant partial integration and the symmetry of $\delta g^{\mu \nu}$ were used.
The contribution of term $\textsc{\romannumeral1}$ to the modified Einstein equation is thus
\begin{equation*}
T^{\textsc{\romannumeral1}}_{\mu\nu} = g_{\mu\nu}\nabla_\beta\left [\nabla_\alpha(\Tilde{f}\phi^{,\alpha}) \,\phi^{,\beta}     \right ]-2\phi_{(,\mu}\nabla_{\nu )}\nabla_\alpha(\Tilde{f}\phi^{,\alpha}) + g_{\mu\nu}\nabla_\beta(\Tilde{Z}\phi^{,\beta} ) -2\Tilde{Z}_{(,\mu}\phi_{,\nu)} 
\end{equation*}
Next, let us turn to term $\textsc{\romannumeral2}$. Using $\delta R_{\mu\nu}$ from above, the fact that $\phi^{,\mu}$ is geodesic (\ref{geodesicness}) and the commutator of covariant derivatives acting on a 2-tensor, we can express
\begin{align*}
    2\phi^{,\mu}\phi^{,\nu}\delta R _{\mu\nu } \doteq \phi_{,\alpha}\phi_{,\beta}\Box \delta g^{\alpha\beta}&+\phi^{,\mu}\nabla_{\mu}\left [\phi^{,\nu}\nabla_{\nu} \left (g_{\alpha\beta}\delta g^{\alpha\beta}  \right )-2\phi_{(,\alpha}\nabla_{\beta)} \delta g^{\alpha\beta}\right ] + \phantom{\frac{1}{1}}\\ 
    &+
2\phi^{,\mu}\left ( \phi_{,\nu} R^{\nu}_{\:\alpha\mu\beta} - \phi_{(,\alpha}R_{\beta)\mu} \right )\delta g^{\alpha\beta}\phantom{\frac{1}{1}}
\end{align*}

The second term is in a form ready for covariant partial integration and the second line does not contain derivatives of $\delta g^{\alpha\beta}$. The first and the third term, however, still need some rewriting. 
\clearpage
\noindent
To this end, calculate 
\begin{equation*}
    \phi_{,\alpha}\phi_{,\beta}\Box\delta g^{\alpha\beta} = \Box\left ( \phi_{,\alpha}\phi_{,\beta}\delta g^{\alpha\beta}   \right )-2\nabla_{\mu}\left ( \nabla^{\mu}\left (\phi_{,\alpha}\phi_{,\beta}    \right )\delta g^{\alpha\beta}\right ) + \Box\left (\phi_{,\alpha}\phi_{,\beta}   \right )\delta g^{\alpha\beta}
\end{equation*}
and
\begin{equation*}
    \phi_{(,\alpha}\nabla_{\beta)} \delta g^{\alpha\beta} = \nabla_{(\alpha}\left ( \phi_{,\beta)} \delta g^{\alpha\beta}   \right )- \phi_{;\alpha\beta} \delta g^{\alpha\beta}. \phantom{\frac{1}{1}}
\end{equation*}
Thus, in summary, we find that
\begin{align*}
2\phi^{,\mu}\phi^{,\nu}\delta R _{\mu\nu } & \doteq  \Box\left ( \phi_{,\alpha}\phi_{,\beta}\delta g^{\alpha\beta}   \right )+\phi^{,\mu}\nabla_{\mu}\left [\phi^{,\nu}\nabla_{\nu} \left (g_{\alpha\beta}\delta g^{\alpha\beta}  \right )-2\nabla_{(\alpha}\left ( \phi_{,\beta)} \delta g^{\alpha\beta}   \right )\right]+\phantom{\frac{1}{1}}\\
&-2\nabla_{\mu}\left ( \nabla^{\mu}\left (\phi_{,\alpha}\phi_{,\beta}    \right ) \delta g^{\alpha\beta}\right ) +2\phi^{,\mu}\nabla_\mu\left (\phi_{;\alpha\beta}   \,\delta g^{\alpha\beta}  \right )+ \phantom{\frac{1}{1}}\\ 
    &+\left (2 \phi^{,\mu}\phi_{,\nu} R^{\nu}_{\:\alpha\mu\beta} - 2\phi^{,\mu}\phi_{(,\alpha}R_{\beta)\mu} + \Box\left (\phi_{,\alpha}\phi_{,\beta}   \right )\right ) \delta g^{\alpha\beta}\phantom{\frac{1}{1}}
\end{align*}
Applying covariant partial integration, we find that the contribution to the modified Einstein equation of term $\textsc{\romannumeral2}$ is given by
\begin{align*}
T^{\textsc{\romannumeral2}}_{\alpha\beta} &=  \phi_{,\alpha}\phi_{,\beta}\Box \Tilde{f}+g_{\alpha\beta}\nabla_\nu\left ( \nabla_\mu( \Tilde{f}\phi^{,\mu})\phi^\nu \right ) -2\phi_{(,\alpha}\nabla_{\beta)}\left ( \nabla_\mu ( \Tilde{f} \phi^{,\mu} ) \right )+\phantom{\frac{1}{1}}\\
&+2\nabla^{\mu}\left (\phi_{,\alpha}\phi_{,\beta}    \right ) \Tilde{f}_{,\mu}-2\phi_{;\alpha\beta} \nabla_\mu(\Tilde{f} \phi^{,\mu})+ \phantom{\frac{1}{1}}\\ 
    &+\left ( 2\phi^{,\mu}\phi_{,\nu} R^{\nu}_{\:\alpha\mu\beta} - 2\phi^{,\mu}\phi_{(,\alpha}R_{\beta)\mu} + \Box\left (\phi_{,\alpha}\phi_{,\beta}   \right )\right ) \Tilde{f}\phantom{\frac{1}{1}}
\end{align*}
Note that the second and third term in the first line cancel the two terms containing $\Tilde{f}$ in $T^{\textsc{\romannumeral1}}_{\alpha\beta}$.

Going on to term $\textsc{\romannumeral3}$, calculate
\begin{equation*}
    \delta_g \left(g^{\mu\alpha}g^{\nu\beta}\phi_{;\mu\nu} \phi_{;\alpha\beta} \right) = 2\phi^{;\mu}_{\alpha} \phi_{;\beta\mu}\, \delta g^{\alpha\beta}-2 \phi^{;\mu\nu}\,\delta \Gamma ^{\lambda}_{\mu\nu}  \phi_{,\lambda}
\end{equation*}
Inserting the variation of the connection coefficients from above, the second term becomes
\begin{align*}
    -2 \phi^{;\mu\nu}\,\delta \Gamma ^{\lambda}_{\mu\nu}  \phi_{,\lambda} &=2 \phi^{;\mu}_{\alpha}\,\phi_{,\beta}\nabla_{\mu} \delta g^{\alpha\beta} - \phi_{;\alpha\beta}\,\phi^{,\lambda}\nabla_{\lambda}\left (\delta g^{\alpha\beta} \right )\phantom{\frac{1}{1}} \\
    &= 2\nabla_\mu\left ( \phi^{;\mu}_{\alpha}\,\phi_{,\beta}\delta g^{\alpha\beta} \right ) - 2\nabla_\mu\left ( \phi^{;\mu}_{\alpha}\,\phi_{,\beta}\right ) \delta g^{\alpha\beta} +\phantom{\frac{1}{1}} \\
&\quad -\phi^{,\lambda}\nabla_{\lambda}\left (\phi_{;\alpha\beta}\delta g^{\alpha\beta} \right ) +\phi^{,\lambda}\nabla_{\lambda}\left (\phi_{;\alpha\beta}\right ) \delta g^{\alpha\beta} \phantom{\frac{1}{1}} 
\end{align*}
where in the second step the expression was brought to a form ready for covariant partial integration.
Summarizing, we find that
\begin{align*}
    \delta_g \left(\phi^{;\mu\nu} \phi_{;\mu\nu} \right) &=  2\nabla_\mu\left ( \phi^{;\mu}_{\alpha}\,\phi_{,\beta}\delta g^{\alpha\beta} \right )-\phi^{,\lambda}\nabla_{\lambda}\left (\phi_{;\alpha\beta}\delta g^{\alpha\beta} \right )  +\phantom{\frac{1}{1}} \\
&\quad +\left (\phi^{,\mu}\phi_{;\alpha\beta\mu}-2\phi_{,\beta}\phi^{;\mu}_{\alpha\mu} \right ) \delta g^{\alpha\beta} \phantom{\frac{1}{1}} 
\end{align*}
Applying covariant partial integration, the contribution of term $\textsc{\romannumeral3}$ to the modified Einstein equation is hence
\begin{equation*}
    T^{\textsc{\romannumeral3}}_{\alpha\beta} =  -\nabla^\mu\left(\phi_{\alpha}\phi_{,\beta}\right)\Tilde{f}_{,\mu}+\phi_{;\alpha\beta}\nabla_\mu(\phi^{,\mu}\Tilde{f})
+\left (\phi^{,\mu}\phi_{;\alpha\beta\mu}-\phi_{,\beta}\phi^{;\mu}_{\alpha\mu}-\phi_{,\alpha}\phi^{;\mu}_{\beta\mu} \right ) \Tilde{f}.
\end{equation*}
Using (\ref{geodesicness}) and the commutator of covariant derivatives we can bring the last term into a form more similar to terms in $T^{\textsc{\romannumeral2}}$ as
\begin{equation*}
    \phi^{,\mu}\phi_{;\alpha\beta\mu}-\phi_{,\beta}\phi^{;\mu}_{\alpha\mu}-\phi_{,\alpha}\phi^{;\mu}_{\beta\mu}  =\phi_{\alpha}^{;\mu}\phi_{;\beta\mu}-R^{\nu}_{\alpha\mu\beta}\phi^{,\mu}\phi_{,\nu}-\Box\left ( \phi_{,\alpha}\phi_{,\beta} \right ).
\end{equation*}
Note that the appearing Riemann tensor components can be rewritten purely in terms of covariant derivatives of $\phi$ according to (\ref{Riemannid}).
Combining all our results, we find that the sum of contibutions to the modified Einstein equation is
\begin{align*}
-T^{\textsc{\romannumeral1}}_{\alpha\beta} +  T^{\textsc{\romannumeral2}}_{\alpha\beta} +  T^{\textsc{\romannumeral3}}_{\alpha\beta} &=  \phi_{,\alpha}\phi_{,\beta}\Box \Tilde{f} -  \nabla_{\mu}(\Tilde{Z}\phi^{,\mu}   )g_{\alpha\beta }  
+2 \phi_{(,\alpha} \Tilde{Z}_{,\beta )}+\phantom{\frac{1}{1}}\\
&+\nabla^{\mu}\left (\phi_{,\alpha}\phi_{,\beta}    \right ) \tilde{f}_{,\mu}- \nabla_\mu(\phi_{;\alpha\beta}\tilde{f} \phi^{,\mu})- 2\phi^{,\mu}\phi_{(,\alpha}R_{\beta)\mu}\tilde{f}\,.\phantom{\frac{1}{1}}
\end{align*}
Finally, term $\textsc{\romannumeral4}$ is easily found to be given by
\begin{equation*}
    T^{\textsc{\romannumeral4}}_{\alpha\beta} = \left(g_{\alpha\beta}\Box-\nabla_\alpha\nabla_\beta+R_{\alpha\beta}\right)h'.
\end{equation*}

\bigskip

\bigskip

\textbf{{\large {Acknowledgments}}}

The work of A. H. C is supported in part by the National Science Foundation
Grant No. Phys-1518371 and Phys-5912998. The work of V.M. and T.B.R. is supported by the Deutsche
Forschungsgemeinschaft (DFG, German Research Foundation) under Germany's
Excellence Strategy -- EXC-2111 -- 390814868. V.M. is grateful to Korea
Institute for Advanced Study, where the part of this work was completed, for hospitality.

\end{document}